\documentclass[letterpaper,twocolumn,10pt]{IEEEtran-cr}

\usepackage{epsfig,endnotes}
\usepackage[utf8]{inputenc}
\usepackage{multirow}
\usepackage{color}
\usepackage[table]{xcolor}
\usepackage[nomessages]{fp}
\usepackage{adjustbox}
\let\labelindent\relax
\usepackage{enumitem}
\usepackage{pgfplots}
\usepackage{balance}

\newcommand{\maxnum}{100.00}
\newlength{\maxlen}
\newcommand{\databar}[2][green!40]{%
  \settowidth{\maxlen}{\maxnum}%
  \addtolength{\maxlen}{\tabcolsep}%
  \FPeval\result{round(#2/\maxnum:4)}%
  \rlap{\color{green!40}\hspace*{-.5\tabcolsep}\rule[-.05\ht\strutbox]{\result\maxlen}{.95\ht\strutbox}}%
  \makebox[\dimexpr\maxlen-\tabcolsep][r]{#2}%
}

\newcommand\eg{\emph{e.g.},\xspace}
\newcommand\ie{\emph{i.e.},\xspace}
\newcommand\etc{\emph{etc}.\xspace}

\providecommand{\etal}{\emph{et al.}\xspace}

\newcommand\nv[1]{\ifstatus\textcolor{green}{}\fi}

\newcommand\claim[1]{\ifstatus\textcolor{black}{#1}\fi}
\newcommand\removed[1]{\ifstatus\textcolor{red}{}\fi}

\newif\ifstatus
\statustrue

\usepackage{graphicx}
\usepackage{subfigure}
\usepackage{xspace}
\DeclareGraphicsExtensions{.png,.eps,.ps,.pdf}

\usepackage{booktabs}
\usepackage{caption}

\usepackage[hyphens]{url}

\usepackage[english]{babel}

\usepackage{xcolor}

\newcommand{\specialcell}[2][c]{%
  \begin{tabular}[#1]{@{}c@{}}#2\end{tabular}}

\newcommand\lumenname{Lumen\xspace}
\newcommand\appname{Firmware Scanner\xspace}

\def\ispFBApps{30\xspace}
\def\nbVendorsMediatek{17\xspace}

\def\appsLibsCustomPerms{94\xspace}

\def\uniqueVendorsGB{6\xspace}
\def\nbVendorsTinno{6\xspace}
\def\androidCertsTW{16\%\xspace}

\def\vendorsShareNL{8\%\xspace}

\def\uniqueVendorsTH{12\xspace}

\def\totalVendorsGB{19\xspace}

\def\totalVendorsTH{28\xspace}
\def\usersAsia{24\%\xspace}
\def\totalLumenAppVersions{298,412\xspace}

\def\fgsWithFacebook{748\xspace}

\def\vendorsShareIT{7\%\xspace}
\def\totalCerts{1,200\xspace}

\def\vendorsShareID{12\%\xspace}

\def\totalVendorsDE{21\xspace}

\def\uniqueVendorsES{3\xspace}

\def\manufacturerCerts{740\xspace}
\def\libAnalyticsNbApps{6,935\xspace}
\def\totalVendorsIT{15\xspace}
\def\fbAppsOnSamsung{68\%\xspace}
\def\usersEurope{35\%\xspace}
\def\uniqueVendorsMX{3\xspace}

\def\totalVendorsID{26\xspace}

\def\vendorsShareDE{10\%\xspace}

\def\libsSamsungCustomPerms{53\%\xspace}

\def\appsNeverUpdatedPerc{74\%\xspace}

\def\vendorsShareUS{17\%\xspace}
\def\androidCertsCN{13\%\xspace}
\def\androidCertsCH{13\%\xspace}

\def\nbVendorsVerizon{5\xspace}

\def\domainsFromPreinstalledInLumen{54,614\xspace}

\def\uniqueVendorsDE{2\xspace}
\def\libSocialNbLibs{70\xspace}

\def\libsIded{334\xspace}

\def\libAdvertisementNbLibsGrouped{107\xspace}

\def\vendorsShareGB{9\%\xspace}
\def\androidCertsUS{43\%\xspace}

\def\totalVendorsNL{16\xspace}

\def\uniqueVendorsUS{11\xspace}
\def\totalLumenApps{139,665\xspace}
\def\appsNeverUpInFiveYearsOrMore{41\%\xspace}

\def\androidCerts{115\xspace}
\def\libSocialNbLibsGrouped{20\xspace}

\def\appsWithFacebook{806\xspace}

\def\appsWithLibs{25,333\xspace}

\def\vendorsShareBE{8\%\xspace}

\def\nbVendorsAeon{3\xspace}

\def\vendorsShareTH{13\%\xspace}
\def\uniqueVendorsBE{4\xspace}
\def\vendorsShareES{11\%\xspace}

\def\rootedDevices{321\xspace}

\def\libAdvertisementNbApps{11,935\xspace}

\def\libAdvertisementNbLibs{164\xspace}
\def\vendorsShareMX{8\%\xspace}

\def\domainsATSFromPreinstalledInLumen{7,629\xspace}
\def\noActNoSrvNoRcv{3,964\xspace}

\def\otherCerts{460\xspace}

\def\totalVendorsMX{17\xspace}

\def\nbVendorsUnknown{1\xspace}

\def\totalVendorsUS{36\xspace}

\def\totalVendorsES{24\xspace}
\def\totalLumenFlows{34,553,193\xspace}
\def\pkgOnGPlay{9\%\xspace}

\def\libSocialNbApps{6,652\xspace}

\def\uniqueVendorsNL{2\xspace}
\def\vendorFBApps{293\xspace}
\def\libsAll{11,665\xspace}

\def\libsCustomPerms{88\xspace}

\def\libAnalyticsNbLibsGrouped{54\xspace}

\def\appsWithInternet{3,118\xspace}
\def\usersAmerica{29\%\xspace}
\def\uniqueVendorsID{7\xspace}

\def\libAnalyticsNbLibs{100\xspace}
\def\totalVendorsBE{17\xspace}
\def\FBApps{98\xspace}
\def\appsInLumenFlows{1,055\xspace}

\def\fileslib64{91,509\xspace}

\def\c10models{1\xspace}
\def\f103_progcp{No\xspace}

\def\f103_prodex{169\xspace}

\def\verizon_lgelibs{11\xspace}

\def\htt:4.2.2dex{80\xspace}

\def\versionP-beta{0.06\%\xspace}
\def\axioo_5.1dex{18\xspace}

\def\cavion7.1quadgcp{No\xspace}

\def\cavion7.1quadfiles{519\xspace}
\def\hi6210sftmodels{1\xspace}

\def\cherry_mobilemodels{1\xspace}

\def\791dex{0\xspace}

\def\c10dex{2\xspace}

\def\p9+certs{162\xspace}

\def\cavion7.1quadmodelsperc{0.06\%\xspace}

\def\jensor2models{1\xspace}

\def\image_mobiledex{5\xspace}

\def\axioo_5.1models{1\xspace}

\def\elite_4.5mgcp{No\xspace}

\def\oos_tcs5056a_advan_s5e_nxt_v1.6_20161223gcp{No\xspace}

\def\elite_4.5mdex{90\xspace}

\def\t36b_egogcp{No\xspace}

\def\t36b_egocerts{150\xspace}

\def\elite_4.5mmodelsperc{0.06\%\xspace}

\def\basco_m500_3ggcp{No\xspace}
\def\alpsapps{2,883\xspace}

\def\histbandroidv5models{1\xspace}

\def\evercoss_a74adex{3\xspace}

\def\motorolaapps{2,158\xspace}

\def\huaweigcp{Yes\xspace}

\def\adm8000kpgcp{No\xspace}

\def\ktc_enmodelsperc{0.06\%\xspace}

\def\adm8000kpcerts{139\xspace}

\def\tm800aapps{30\xspace}

\def\mrvl-mgmodels{1\xspace}

\def\c10libs{6\xspace}
\def\rk2928sdklibs{70\xspace}

\def\t36b_egoapps{111\xspace}

\def\lmy47dmodelsperc{0.06\%\xspace}

\def\verizon_lgeapps{0\xspace}
\def\v403libs{266\xspace}

\def\4gdex{0\xspace}

\def\image_mobilegcp{No\xspace}

\def\htt:4.2.2gcp{No\xspace}

\def\adm8000kplibs{197\xspace}

\def\zh960files{812\xspace}

\def\jensor2libs{257\xspace}

\def\locationid{6\%\xspace}

\def\verizon_lgegcp{No\xspace}

\def\elite_4.5mcerts{162\xspace}

\def\axioo_5.1files{908\xspace}
\def\condor_electronicsdex{1\xspace}

\def\htc_europemodelsperc{0.06\%\xspace}

\def\jensor2apps{72\xspace}
\def\locationit{5\%\xspace}

\def\evercoss_a74agcp{No\xspace}

\def\sf1certs{157\xspace}

\def\tm800acerts{158\xspace}
\def\htt:4.2.2libs{378\xspace}

\def\mt6577modelsperc{0.06\%\xspace}

\def\htc_europecerts{176\xspace}

\def\rk30sdkdex{0\xspace}

\def\oos_tcs5056a_advan_s5e_nxt_v1.6_20161223dex{12\xspace}

\def\locationgb{4\%\xspace}

\def\totaldevices{1,742\xspace}

\def\cherry_mobileapps{15\xspace}

\def\oos_tcs5056a_advan_s5e_nxt_v1.6_20161223apps{36\xspace}

\def\locationus{12\%\xspace}

\def\zh960libs{422\xspace}

\def\condor_electronicsapps{10\xspace}

\def\histbandroidv5apps{74\xspace}

\def\itab-703scerts{140\xspace}
\def\ktc_endex{0\xspace}

\def\hi6210sftdex{35\xspace}

\def\devicesGPtrotect{22\%\xspace}

\def\4gfiles{879\xspace}

\def\mt6577models{1\xspace}

\def\ktc_engcp{No\xspace}

\def\verizon_lgedex{0\xspace}
\def\rk2928sdkdex{0\xspace}

\def\locationth{3\%\xspace}
\def\zh960dex{4\xspace}

\def\image_mobilelibs{33\xspace}

\def\tm800adex{19\xspace}

\def\rk30sdkgcp{No\xspace}

\def\rk2928sdkgcp{No\xspace}
\def\f103_prolibs{994\xspace}

\def\rk2928sdkcerts{139\xspace}

\def\topwise-tabletmodels{1\xspace}
\def\inet-tabletgcp{No\xspace}

\def\791files{559\xspace}

\def\topwise-tabletlibs{26\xspace}

\def\totalvendors{214\xspace}

\def\samsungmodelsperc{25\%\xspace}

\def\m4telmodelsperc{0.06\%\xspace}

\def\itab-703slibs{282\xspace}

\def\axioo_5.1certs{162\xspace}

\def\zh960models{1\xspace}

\def\basco_m500_3gdex{0\xspace}

\def\cherry_mobilefiles{889\xspace}

\def\axioo_5.1apps{37\xspace}
\def\locationes{6\%\xspace}

\def\jensor2gcp{No\xspace}
\def\mrvl-mgmodelsperc{0.06\%\xspace}

\def\evercoss_a74alibs{290\xspace}

\def\cherry_mobilelibs{34\xspace}
\def\mrvl-mgfiles{639\xspace}

\def\inet-tabletdex{0\xspace}

\def\tm800amodelsperc{0.06\%\xspace}

\def\m4telmodels{1\xspace}

\def\mt6577dex{66\xspace}

\def\mt6577gcp{No\xspace}

\def\rk30sdkcerts{140\xspace}
\def\ktc_enapps{63\xspace}

\def\axioo_5.1modelsperc{0.06\%\xspace}
\def\histbandroidv5gcp{No\xspace}

\def\c10files{17\xspace}
\def\htc_europegcp{No\xspace}
\def\samsungfiles{260,187\xspace}

\def\p9+dex{111\xspace}

\def\htc_europedex{253\xspace}

\def\evercoss_a74afiles{819\xspace}
\def\791apps{20\xspace}

\def\cavion7.1quadcerts{156\xspace}

\def\p9+modelsperc{0.06\%\xspace}

\def\basco_m500_3gfiles{848\xspace}

\def\locationde{3\%\xspace}

\def\htc_europemodels{1\xspace}

\def\tm800agcp{No\xspace}

\def\tm800amodels{1\xspace}

\def\lgeapps{3,596\xspace}

\def\sf1modelsperc{0.06\%\xspace}

\def\lmy47dcerts{162\xspace}
\def\rk2928sdkmodels{2\xspace}

\def\sf1models{1\xspace}
\def\inet-tabletlibs{20\xspace}
\def\lmy47dapps{103\xspace}

\def\m4telcerts{0\xspace}

\def\hi6210sftcerts{158\xspace}

\def\htt:4.2.2apps{120\xspace}

\def\condor_electronicsmodelsperc{0.06\%\xspace}

\def\elite_4.5mlibs{427\xspace}

\def\m4teldex{0\xspace}

\def\evercoss_a74amodels{1\xspace}

\def\inet-tabletmodelsperc{0.06\%\xspace}

\def\oos_tcs5056a_advan_s5e_nxt_v1.6_20161223certs{162\xspace}

\def\lmy47dfiles{708\xspace}

\def\f103_profiles{1,516\xspace}

\def\791models{1\xspace}
\def\inet-tabletmodels{1\xspace}
\def\zh960gcp{No\xspace}

\def\zh960modelsperc{0.06\%\xspace}
\def\f103_promodels{1\xspace}

\def\791certs{162\xspace}

\def\lmy47dgcp{No\xspace}

\def\fileslib{184,362\xspace}

\def\cavion7.1quadmodels{1\xspace}

\def\cherry_mobilemodelsperc{0.06\%\xspace}

\def\sf1gcp{No\xspace}

\def\htt:4.2.2certs{140\xspace}

\def\m4tellibs{0\xspace}

\def\oos_tcs5056a_advan_s5e_nxt_v1.6_20161223files{745\xspace}
\def\htt:4.2.2modelsperc{0.06\%\xspace}

\def\t36b_egolibs{450\xspace}

\def\v403models{1\xspace}

\def\histbandroidv5files{695\xspace}

\def\locationbe{2\%\xspace}

\def\791modelsperc{0.06\%\xspace}

\def\mrvl-mgapps{11\xspace}

\def\htt:4.2.2models{1\xspace}

\def\f103_promodelsperc{0.06\%\xspace}

\def\evercoss_a74acerts{162\xspace}
\def\v403gcp{No\xspace}

\def\topwise-tabletfiles{477\xspace}

\def\evercoss_a74aapps{24\xspace}

\def\histbandroidv5certs{168\xspace}

\def\motorolafiles{28,291\xspace}

\def\basco_m500_3gcerts{162\xspace}

\def\cherry_mobilecerts{158\xspace}
\def\hi6210sftmodelsperc{0.06\%\xspace}

\def\m4telgcp{Yes\xspace}

\def\htc_europelibs{407\xspace}

\def\ktc_encerts{146\xspace}

\def\t36b_egodex{80\xspace}
\def\jensor2certs{140\xspace}

\def\adm8000kpmodels{1\xspace}

\def\basco_m500_3gmodels{1\xspace}

\def\axioo_5.1libs{386\xspace}

\def\jensor2files{489\xspace}

\def\oos_tcs5056a_advan_s5e_nxt_v1.6_20161223modelsperc{0.06\%\xspace}

\def\rk30sdkmodelsperc{0.06\%\xspace}

\def\image_mobilemodelsperc{0.06\%\xspace}
\def\htc_europeapps{259\xspace}
\def\inet-tabletfiles{36\xspace}

\def\verizon_lgemodelsperc{0.06\%\xspace}

\def\f103_proapps{181\xspace}
\def\rk2928sdkapps{33\xspace}

\def\image_mobileapps{17\xspace}

\def\itab-703sfiles{722\xspace}

\def\jensor2dex{0\xspace}

\def\sf1files{2,131\xspace}

\def\rk30sdklibs{273\xspace}

\def\791libs{64\xspace}

\def\hi6210sftlibs{858\xspace}
\def\rk30sdkmodels{1\xspace}
\def\topwise-tabletapps{22\xspace}

\def\oos_tcs5056a_advan_s5e_nxt_v1.6_20161223libs{289\xspace}

\def\topwise-tabletdex{0\xspace}

\def\v403modelsperc{0.06\%\xspace}

\def\hi6210sftfiles{1,190\xspace}

\def\hi6210sftapps{127\xspace}
\def\inet-tabletcerts{7\xspace}

\def\noActNoSrvNoRcv{5,244\xspace}

\def\rk30sdkapps{90\xspace}

\def\adm8000kpapps{62\xspace}
\def\cherry_mobiledex{0\xspace}

\def\totalapps{82,501\xspace}
\def\t36b_egofiles{816\xspace}

\def\evercoss_a74amodelsperc{0.06\%\xspace}

\def\mt6577files{678\xspace}

\def\lgefiles{58,273\xspace}

\def\topwise-tabletgcp{No\xspace}

\def\ktc_enmodels{1\xspace}
\def\condor_electronicsfiles{686\xspace}

\def\totalusers{2,748\xspace}

\def\topwise-tabletcerts{140\xspace}

\def\c10modelsperc{0.06\%\xspace}
\def\m4telfiles{0\xspace}

\def\4gcerts{158\xspace}

\def\condor_electronicsmodels{1\xspace}

\def\verizon_lgecerts{8\xspace}

\def\hi6210sftgcp{No\xspace}

\def\sf1apps{26\xspace}
\def\huaweiapps{12,401\xspace}
\def\4glibs{44\xspace}

\def\axioo_5.1gcp{No\xspace}

\def\verizon_lgefiles{53\xspace}

\def\mt6577libs{380\xspace}
\def\totalfiles{424,584\xspace}

\def\histbandroidv5modelsperc{0.06\%\xspace}

\def\adm8000kpfiles{503\xspace}

\def\p9+models{1\xspace}

\def\791gcp{No\xspace}
\def\huaweimodelsperc{20\%\xspace}

\def\oos_tcs5056a_advan_s5e_nxt_v1.6_20161223models{1\xspace}
\def\zh960certs{170\xspace}

\def\image_mobilefiles{731\xspace}
\def\ktc_enlibs{375\xspace}

\def\elite_4.5mmodels{1\xspace}

\def\4gmodels{1\xspace}

\def\condor_electronicscerts{140\xspace}

\def\p9+libs{506\xspace}

\def\elite_4.5mapps{102\xspace}
\def\m4telapps{0\xspace}

\def\f103_procerts{158\xspace}
\def\cavion7.1quaddex{0\xspace}

\def\c10apps{2\xspace}

\def\htt:4.2.2files{742\xspace}

\def\tm800alibs{81\xspace}

\def\image_mobilemodels{1\xspace}

\def\c10gcp{No\xspace}

\def\condor_electronicsgcp{No\xspace}

\def\v403files{746\xspace}

\def\lmy47dlibs{336\xspace}
\def\basco_m500_3gapps{9\xspace}

\def\t36b_egomodels{1\xspace}

\def\lmy47ddex{97\xspace}

\def\condor_electronicslibs{71\xspace}
\def\cherry_mobilegcp{Yes\xspace}
\def\huaweifiles{150,405\xspace}

\def\histbandroidv5dex{0\xspace}

\def\mrvl-mgdex{1\xspace}

\def\basco_m500_3glibs{374\xspace}

\def\histbandroidv5libs{444\xspace}

\def\p9+files{904\xspace}

\def\topwise-tabletmodelsperc{0.06\%\xspace}
\def\adm8000kpdex{37\xspace}

\def\mrvl-mggcp{No\xspace}

\def\4ggcp{No\xspace}

\def\ktc_enfiles{592\xspace}
\def\tm800afiles{837\xspace}

\def\rk2928sdkfiles{738\xspace}

\def\locationnl{2\%\xspace}

\def\mt6577certs{139\xspace}

\def\basco_m500_3gmodelsperc{0.06\%\xspace}

\def\v403dex{3\xspace}

\def\cavion7.1quadapps{27\xspace}

\def\jensor2modelsperc{0.06\%\xspace}
\def\4gmodelsperc{0.06\%\xspace}
\def\mt6577apps{87\xspace}

\def\inet-tabletapps{9\xspace}
\def\alpsgcp{No\xspace}

\def\v403apps{22\xspace}

\def\elite_4.5mfiles{927\xspace}

\def\itab-703smodelsperc{0.06\%\xspace}

\def\adm8000kpmodelsperc{0.06\%\xspace}

\def\itab-703smodels{1\xspace}

\def\t36b_egomodelsperc{0.06\%\xspace}

\def\sf1libs{478\xspace}

\def\locationmx{3\%\xspace}

\def\mrvl-mglibs{50\xspace}

\def\lmy47dmodels{1\xspace}

\def\cavion7.1quadlibs{46\xspace}

\def\locationsall{130\xspace}

\def\image_mobilecerts{158\xspace}

\def\verizon_lgemodels{1\xspace}

\def\v403certs{162\xspace}

\def\rk2928sdkmodelsperc{0.11\%\xspace}
\def\rk30sdkfiles{526\xspace}

\def\htc_europefiles{1,101\xspace}

\def\samsungapps{29,466\xspace}

\def\itab-703sdex{8\xspace}

\def\zh960apps{35\xspace}
\def\sf1dex{0\xspace}

\def\4gapps{17\xspace}

\def\p9+gcp{No\xspace}

\def\itab-703sapps{36\xspace}

\def\c10certs{7\xspace}

\def\mrvl-mgcerts{134\xspace}
\def\itab-703sgcp{No\xspace}

\def\samsunggcp{Yes\xspace}

\def\p9+apps{112\xspace}

\def\alpsfiles{29,288\xspace}

\def\totalLumenCountries{144\xspace}
\def\totalLumenVendors{291\xspace}
\def\totalLumenUsers{20.4K\xspace}

\def\devsmotorola{50\xspace}
\def\devsalps{65\xspace}
\def\devslge{74\xspace}
\def\devshuawei{343\xspace}
\def\devssamsung{441\xspace}

\def\usersmotorola{110\xspace}
\def\usersalps{136\xspace}
\def\userslge{154\xspace}
\def\usershuawei{716\xspace}
\def\userssamsung{924\xspace}

\def\filesmotorola{801\xspace}
\def\appsmotorola{127\xspace}
\def\libsmotorola{454\xspace}
\def\dexsmotorola{62\xspace}
\def\certsmotorola{151\xspace}
\def\filesalps{632\xspace}
\def\appsalps{56\xspace}
\def\libsalps{385\xspace}
\def\dexsalps{46\xspace}
\def\certsalps{148\xspace}
\def\fileslge{675\xspace}
\def\appslge{84\xspace}
\def\libslge{385\xspace}
\def\dexslge{89\xspace}
\def\certslge{150\xspace}
\def\fileshuawei{1,084\xspace}
\def\appshuawei{68\xspace}
\def\libshuawei{766\xspace}
\def\dexshuawei{96\xspace}
\def\certshuawei{146\xspace}
\def\filessamsung{868\xspace}
\def\appssamsung{136\xspace}
\def\libssamsung{556\xspace}
\def\dexssamsung{83\xspace}
\def\certssamsung{150\xspace}

\def\totalNonOfficialMNOTotalVendors{15\xspace}
\def\totalPackagesDeclarigCustomPerms{1,795\xspace}
\def\totalNonOfficialVendorTotalPerms{3,760\xspace}
\def\totalNonOfficialOtherTotalPerms{549\xspace}
\def\totalNonOfficialbrowserTotalVendors{6\xspace}
\def\totalNonOfficialchipsetTotalPerms{67\xspace}

\def\percentageSamsungVendorTotalPerms{41\%\xspace}
\def\percentageSonyVendorTotalPerms{7\%\xspace}

\def\totalNonOfficialAllianceTotalVendors{44\xspace}
\def\totalNonOfficialchipsetTotalVendors{63\xspace}
\def\totalNonOfficialAllianceTotalPerms{29\xspace}

\def\totalNonOfficialMNOTotalPerms{195\xspace}

\def\totalNonOfficialavSecTotalPerms{46\xspace}

\def\percentageHuaweiVendorTotalPerms{20\%\xspace}
\def\totalVendorsDeclarigCustomPerms{108\xspace}

\def\totalNonOfficialVendorTotalProviders{31\xspace}

\def\totalNonOfficialVendorTotalVendors{37\xspace}

\def\totalNonOfficialPermsWithAOSPrefix{269\xspace}
\def\totalNonOfficialbrowserTotalPerms{7\xspace}

\def\totalNonOfficialtpTotalVendors{34\xspace}
\def\totalNonOfficialtpTotalPerms{192\xspace}
\def\totalNonOfficialavSecTotalVendors{13\xspace}
\def\percentageVendorTotalPerms{63\%\xspace}
\def\totalCustomPerms{4,845\xspace}

\def\totalNonOfficialOtherTotalVendors{75\xspace}

\def\percentageUnclassifiedPerms{9\%\xspace}

\def\totalAndroidCustomPerms{494\xspace}

\def\totalAndroidSettingsCustomPermsVendors{16\xspace}
\def\totalAndroidEmailCustomPerms{33\xspace}
\def\totalAndroidMMSCustomPerms{59\xspace}
\def\totalAndroidEmailCustomPermsVendors{10\xspace}

\def\totalAndroidCustomPermsVendors{21\xspace}

\def\totalAndroidMSSCustomPermsVendors{11\xspace}
\def\totalAndroidPhoneCustomPerms{84\xspace}
\def\totalAndroidUICustomPermsVendors{15\xspace}

\def\totalAndroidContactsCustomPerms{40\xspace}
\def\totalAndroidUICustomPerms{90\xspace}
\def\totalAndroidPhoneCustomPermsVendors{14\xspace}

\def\totalVendorsCustomPerms{108\xspace}
\def\totalAndroidContactsCustomPermsVendors{7\xspace}
\def\totalCustomPerms{4,845\xspace}
\def\totalAndroidSettingsCustomPerms{87\xspace}

\def\totalAndroidPhoneVendorCustomPerms{56\xspace}
\def\totalAndroidVendorCustomPermsVendors{9\xspace}
\def\totalAndroidContactsVendorCustomPermsVendors{3\xspace}

\def\totalAndroidPhoneVendorCustomPermsVendors{9\xspace}
\def\totalAndroidUIVendorCustomPermsVendors{11\xspace}

\def\totalAndroidEmailVendorCustomPerms{18\xspace}
\def\totalAndroidMMSVendorCustomPermsVendors{10\xspace}
\def\totalAndroidEmailVendorCustomPermsVendors{4\xspace}
\def\totalAndroidContactsVendorCustomPerms{32\xspace}
\def\totalAndroidVendorCustomPerms{410\xspace}
\def\totalAndroidSettingsVendorCustomPerms{63\xspace}

\def\totalAndroidMMSVendorCustomPerms{35\xspace}
\def\totalAndroidUIVendorCustomPerms{67\xspace}

\def\totalAndroidSettingsVendorCustomPermsVendors{12\xspace}

\def\totalAndroidMMSMNOCustomPerms{1\xspace}
\def\totalAndroidMNOCustomPermsVendors{2\xspace}
\def\totalAndroidPhoneMNOCustomPerms{5\xspace}
\def\totalAndroidMNOCustomPerms{12\xspace}
\def\totalAndroidSettingsMNOCustomPermsVendors{1\xspace}
\def\totalAndroidMMSMNOCustomPermsVendors{2\xspace}
\def\totalAndroidSettingsMNOCustomPerms{1\xspace}
\def\totalAndroidPhoneMNOCustomPermsVendors{2\xspace}

\def\totalAndroidUIThirdPCustomPerms{1\xspace}
\def\customcomfacebooksystemVendors{18\xspace}
\def\customcomfacebookorcaPermissions{5\xspace}
\def\customcomfacebookappmanagerVendors{15\xspace}
\def\totalDTPermissionsCustomPerms{8\xspace}

\def\totalFBPerms{18\xspace}

\def\customcomfacebookappmanagerPermissions{4\xspace}

\def\customcomfacebookorcaVendors{5\xspace}

\def\customcomfacebooksystemPermissions{2\xspace}
\def\totalFBVendors{24\xspace}
\def\customcomfacebookpagesappVendors{1\xspace}

\def\customcomfacebookpagesappPermissions{4\xspace}

\def\customcomfacebookliteVendors{1\xspace}
\def\customcomfacebooklitePermissions{1\xspace}

\def\totalAndroidUIThirdPCustomPermsVendors{2\xspace}
\def\totalBaiduVendorsCustomPerms{7\xspace}
\def\customcomfacebookkatanaPermissions{8\xspace}
\def\customcomfacebookkatanaVendors{14\xspace}

\def\totalAndroidMMSAllianceCustomPerms{1\xspace}
\def\totalAndroidMMSAllianceCustomPermsVendors{1\xspace}
\def\totalAndroidAllianceCustomPermsVendors{7\xspace}

\def\totalAndroidAllianceCustomPerms{6\xspace}

\def\totalAndroidPhoneChipsetCustomPermsVendors{5\xspace}
\def\totalAndroidChipsetCustomPerms{4\xspace}
\def\totalAndroidPhoneChipsetCustomPerms{3\xspace}
\def\totalAndroidChipsetCustomPermsVendors{13\xspace}

\def\totalAndroidPhoneOTHERCustomPerms{20\xspace}

\def\totalAndroidOTHERCustomPerms{62\xspace}
\def\totalAndroidContactsOTHERCustomPermsVendors{5\xspace}
\def\totalAndroidMMSOTHERCustomPermsVendors{8\xspace}
\def\totalAndroidContactsOTHERCustomPerms{8\xspace}
\def\totalAndroidPhoneOTHERCustomPermsVendors{10\xspace}
\def\totalAndroidOTHERCustomPermsVendors{17\xspace}

\def\totalAndroidUIOTHERCustomPermsVendors{8\xspace}

\def\totalAndroidSettingsOTHERCustomPermsVendors{8\xspace}

\def\totalAndroidEmailOTHERCustomPerms{15\xspace}
\def\totalAndroidMMSOTHERCustomPerms{22\xspace}
\def\totalAndroidUIOTHERCustomPerms{22\xspace}
\def\totalAndroidSettingsOTHERCustomPerms{23\xspace}

\def\totalAppsUsingPerms{4,736\xspace}
\def\totalAppsUsingHundredPerms{55\xspace}

\begin{document}
\hyphenation{An-dro-Sniff}

\date{}

\title{\Large \bf An Analysis of Pre-installed Android Software}

 \author{
   \IEEEauthorblockN{Julien Gamba\IEEEauthorrefmark{1}\IEEEauthorrefmark{2},
     Mohammed Rashed\IEEEauthorrefmark{2},
     Abbas Razaghpanah\IEEEauthorrefmark{3},
     Juan Tapiador \IEEEauthorrefmark{2} and
     Narseo Vallina-Rodriguez\IEEEauthorrefmark{1}\IEEEauthorrefmark{4}} \\ \vspace{2mm}
   \IEEEauthorblockA{\IEEEauthorrefmark{1} IMDEA Networks Institute, }
   \IEEEauthorblockA{\IEEEauthorrefmark{2} Universidad Carlos III de Madrid,}
   \IEEEauthorblockA{\IEEEauthorrefmark{3} Stony Brook University, } 
   \IEEEauthorblockA{\IEEEauthorrefmark{4} ICSI}
} 

\maketitle

\subsection*{Abstract}

The open-source nature of the Android OS makes it possible for manufacturers to 
ship custom versions of the OS along with a set of pre-installed apps, often for 
product differentiation. Some device vendors have recently come under scrutiny 
for potentially invasive private data collection practices and other
potentially harmful or unwanted behavior of
the pre-installed apps on their devices. Yet, the landscape of pre-installed 
software in Android has largely remained unexplored, particularly in terms of the 
security and privacy implications of such customizations. In this paper, we present 
the first large-scale study of pre-installed software on Android devices from
more than 200 
vendors. Our work relies on a large dataset of real-world Android firmware acquired 
worldwide using crowd-sourcing methods. This allows us to answer questions related 
to the stakeholders involved in the supply chain, from device manufacturers and mobile 
network operators to third-party organizations like advertising and tracking
services, 
and social network platforms. Our study allows us to also uncover 
relationships between these actors, which seem to revolve primarily around advertising and 
data-driven services. Overall, the supply chain around Android's 
open source model lacks transparency and 
has facilitated potentially harmful behaviors and 
backdoored access to sensitive data and services without user consent or
awareness. We conclude the paper with recommendations to improve transparency, 
attribution, and accountability in the Android ecosystem.

\vspace*{-.5em}
\section{Introduction}
\label{sec:intro}
The openness of the Android source code makes it possible for any manufacturer to
ship a custom version of the OS along with proprietary pre-installed 
apps on the system partition. Most handset vendors take this opportunity to add value 
to their products as a market differentiator, typically through partnerships with 
Mobile Network Operators (MNOs), online social networks, and content providers. 
Google does not forbid this behavior, and it has developed its Android Compatibility
Program~\cite{androidcompat} to set the requirements that the modified OS must
fulfill in order to remain compatible with standard Android apps, regardless of
the modifications introduced.
Devices made by vendors that are part of the Android Certified Partners 
program~\cite{androidcertification} come pre-loaded with Google's suite of 
apps (\eg the Play Store and Youtube). Google does not provide details about 
the certification processes. Companies that want to include the Google Play 
service without the certification can outsource the design of the product to 
a certified Original Design Manufacturer (ODM)~\cite{certifiedpartners}.

Certified or not, not all pre-installed software is deemed as wanted by users, and 
the term ``bloatware'' is often applied to such software. The process of 
how a particular set of apps end up packaged together in the firmware of a device 
is not transparent, and various isolated  
cases reported over the last few years suggest that
it lacks end-to-end control mechanisms to guarantee that shipped firmware is
free from vulnerabilities~\cite{huaweiGameSkytoneCVE,samsungAccountCVE} or
potentially malicious 
and unwanted apps. For example, at Black Hat USA 2017, 
Johnson \etal{}~\cite{johnson17bhus,kryptowireADUPSfollowup} gave details of a powerful 
backdoor present in the firmware of several models of Android smartphones, including 
the popular BLU R1 HD.  
In response to this disclosure, Amazon removed Blu products from their 
Prime Exclusive line-up~\cite{amazonBluPhones}. 
A company named Shanghai Adups Technology Co.\ Ltd.\ was pinpointed as responsible for
this incident. The same report also discussed the case of how vulnerable core system 
services (\eg the widely deployed MTKLogger component developed by the chipset
manufacturer MediaTek) could be abused by co-located apps. 
The infamous Triada trojan has also been recently found embedded in the firmware
of several low-cost Android smartphones~\cite{triadaTrojan,triadaLow}. Other cases of 
malware found pre-installed include Loki (spyware and adware) and Slocker (ransomware), 
which were spotted in the firmware of various high-end
phones~\cite{LokiPreinstalled}.

Android handsets also play a key role in the mass-scale data collection practices followed 
by many actors in the digital economy, including advertising and tracking companies. 
OnePlus has been under suspicion of collecting personally identifiable information (PII) from users
of its smartphones through exceedingly detailed analytics~\cite{oneplus,oneplus1}, and also 
deploying the capability to remotely root the phone~\cite{oneplus2,oneplus3}. In July 2018 
the New York Times revealed the existence of secret agreements between Facebook and device 
manufacturers such as Samsung~\cite{facebook} to collect private data from users without
their knowledge. This is currently under investigation by the US Federal
authorities~\cite{nytfbinvestigation}. 
Additionally, users from developing countries with lax data protection and privacy 
laws may be at an even greater risk. The Wall Street Journal has exposed the
presence of a pre-installed app that sends users' geographical location as well as device identifiers 
to GMobi, a mobile-advertising agency that engages in ad-fraud activities~\cite{newley,upstreamSystems}.
Recently, the European Commission publicly expressed concern about Chinese manufacturers like Huawei, 
alleging that they were required to cooperate with national intelligence services by installing 
backdoors on their devices~\cite{huaweieu}.

\vspace*{.5em}
\noindent\textit{Research Goals and Findings}

To the best of our knowledge, no research study has so far systematically
studied the vast ecosystem of pre-installed Android software and the privacy
and security concerns associated with them. This ecosystem has remained largely
unexplored due to the inherent difficulty to access such software at scale and
across vendors. This state of affairs makes such an study even more relevant,
since $i)$ these apps -- typically unavailable on app stores --
have mostly escaped the
scrutiny of researchers and regulators; and $ii)$ regular users are unaware of their
presence on the device, which could imply lack of consent in data collection
and other activities.

In this paper, we seek to shed light on the presence and behavior of pre-installed software 
across Android devices. In particular, we aim to answer the questions below:

\begin{itemize}[leftmargin=*]

  \item What is the ecosystem of pre-installed apps, including all actors in
   the supply chain?

  \item What are the relationships between vendors and other stakeholders
  (\eg{} MNOs and third-party services)?

  \item Do pre-installed apps collect private and personally-identifiable 
  information (PII)? If so, with whom do they share it?

  \item Are there any \claim{harmful} or other potentially dangerous apps among pre-installed
  software?

\end{itemize}

To address the points described above, we developed a research agenda revolving around 
four main items:

\begin{enumerate}[leftmargin=*]

  \item We collected the firmware and traffic information from real-world devices
  using crowd-sourcing methods (\S\ref{sec:dataset}). We obtained the
  firmware from \totalusers{} users spanning \totaldevices{} device models 
  from \totalvendors{} vendors. Our user base covers \locationsall{} countries
  from the main Android markets. Our dataset contains
  \totalfiles{} unique firmware files, but only 9\% of the collected APKs were
  found in Google Play. 
  \claim{We complement this dataset with
  traffic flows associated with \totalLumenApps{} unique 
  apps, including pre-installed ones},
    provided by over \totalLumenUsers{} users of the Lumen app~\cite{lumen}
  from \totalLumenCountries{} countries.  
  To the best of our knowledge, this is the largest dataset of real-world Android 
  firmware analyzed so far.

  \item We performed an investigation of the ecosystem of pre-installed Android
  apps and the actors involved (\S\ref{sec:ecosystem}) by analyzing the 
  Android manifest files of the app packages, their certificates, and the Third-Party 
  Libraries (TPLs) they use. Our analysis covers 1,200 unique developers
  associated 
  with major manufacturers, vendors, MNOs, and Internet service companies. We also uncover a 
  vast landscape of third-party libraries (11,665 unique TPLs), many of which mainly 
  provide data-driven services such as advertisement, analytics, and social networking.

  \item We extracted and analyzed an extensive set of custom permissions
  (4,845) declared by hardware vendors, MNOs, third-party services, security
  firms, industry alliances, chipset manufacturers, and Internet browsers. Such
  permissions may potentially 
  expose data and features to over-the-top apps and could be used to
  access privileged system resources and sensitive data in a way that circumvents
  the Android permission model. A manual inspection reveals a complex supply 
  chain that involves different stakeholders and potential 
  commercial partnerships between them (\S\ref{sec:permissions}).

  \item We carried out a behavioral analysis of nearly 50\% of the apps in our dataset 
  using both static and dynamic analysis tools (\S\ref{sec:appAnalysis}). Our 
  results reveal that a significant part of the pre-installed software exhibit potentially 
  \claim{harmful} or unwanted behavior. \claim{While it is known that personal
  data collection and user tracking is pervasive in the 
  Android app ecosystem as a whole~\cite{felt2011android,pan2018panoptispy,razaghpanah2018apps}, we find that it is also quite prevalent in pre-installed 
  apps}. 
  \claim{We have identified instances of user tracking activities 
  by pre-installed Android software -- and embedded third-party libraries -- 
  which range from collecting the usual set 
  of PII and geolocation data to more invasive practices 
  that include personal email and phone call metadata, contacts, and a variety of behavioral
  and usage statistics in some cases.}
  We also found a few isolated malware samples belonging to known families, according to
  VirusTotal, with prevalence 
  in the last few years (\eg Xynyin, SnowFox, Rootnik, Triada and Ztorg), and generic
  trojans displaying a standard set of malicious behaviors (\eg silent app promotion, SMS 
  fraud, ad fraud, and URL click fraud).

\end{enumerate}

All in all, our work reveals complex relationships between actors in the Android
ecosystem, \claim{in which user data seems to be a major commodity.}
We uncover a myriad of actors involved in the
development of mobile software, as well as \claim{poor software engineering
practices and lack of transparency in the supply chain} that unnecessarily increase
users' security and privacy risks. 
We conclude this paper with various recommendations to palliate this state of
affairs, including transparency models to improve attribution and
accountability, and clearer mechanisms to obtain informed consent. Given the 
scale of the ecosystem and the need to perform manual inspections, we will 
gradually make our dataset 
available to the research community and regulators 
to boost investigations.

\section{Data Collection}
\label{sec:dataset}

Obtaining pre-installed apps and other software artifacts (\eg{} certificates
installed in the system root store) at scale is challenging. As purchasing 
all the mobile handset models (and their many variations) available in the market
is unfeasible, we decided to crowdsource the collection of pre-installed software 
using a purpose-built app: \appname{}~\cite{scannerapp}. Using \appname{}, we obtained 
pre-installed software from \totaldevices{} device models. We also decided to 
use \lumenname{}, an app that aims to promote mobile transparency and enable
user control over their mobile traffic~\cite{lumen,lumenapp} to 
obtain anonymized network flow metadata from \lumenname{}'s real users. 
This allows us to correlate the information we extract from static analysis, for 
a subset of mobile apps, with realistic network traffic generated by mobile 
users in the wild and captured in user-space. 
In the remainder of this section, we explain the methods 
implemented by each app and present our datasets. We discuss the ethical 
implications of our data collection in Section~\ref{sub:ethics}.

\begin{figure*}[t]
  \centering
  \includegraphics[width=0.95\linewidth]{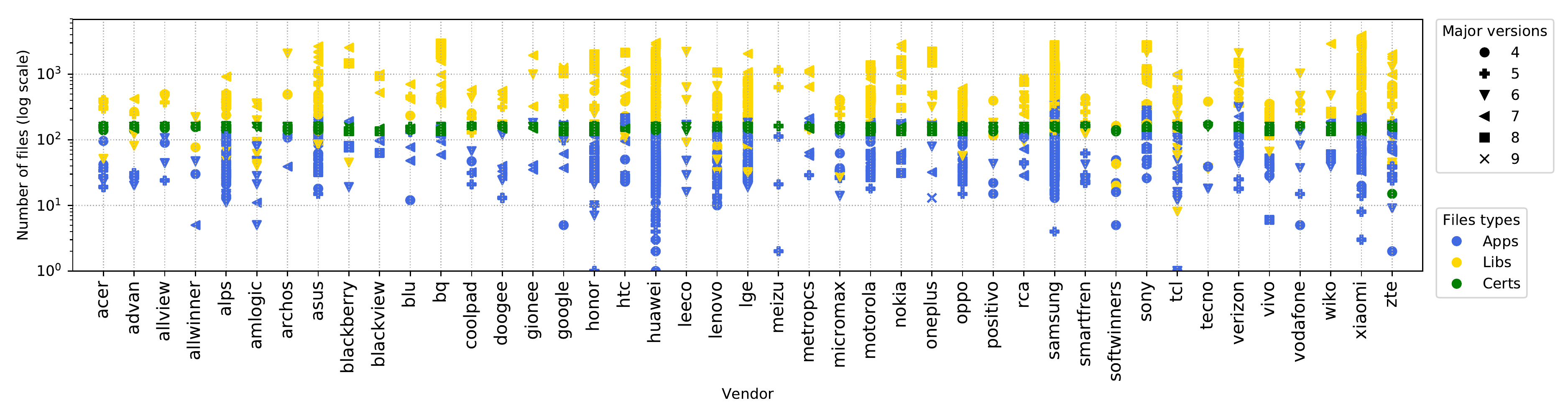}
  \vspace*{-1em}
  \caption{Number of files per vendor. We do not display
  the vendors for which we have less than 3 devices.}
  \vspace*{-1.5em}
\label{fig:nb-files-per-vendor}
\end{figure*}

\vspace*{-1em}
\subsection{\appname{}}
\label{sec:firmwarescanner}

Publicly available on Google Play~\cite{scannerapp}, \appname{} is a 
purpose-built Android app that looks for and extracts pre-installed apps 
and DEX files in the \texttt{app} and \texttt{priv-app} folders located 
in \texttt{/system/}, libraries in the \texttt{lib} and \texttt{lib64} 
folders in \texttt{/system/}, any files in the \texttt{/system/vendor/}
folder if that directory exists, and root certificates located 
in \texttt{/system/etc/security/cacerts/}. We can distinguish pre-installed
apps from user-installed ones as the latter are 
stored in \texttt{/data/app/}. In order to reduce the scanning and 
upload time, \appname{} first computes the MD5 hashes of the relevant files 
(\eg{} apps, libraries, and root certificates) and then sends the list of these 
hashes to our server. Only those missing in our dataset are uploaded over a Wi-Fi 
connection to avoid affecting the user's data plan.

\noindent \textbf{Dataset: } 
Thanks to \totalusers{} users who have organically installed \appname{}, we obtained 
firmware versions for \totaldevices{} unique device models\footnotemark[1] branded by
\totalvendors{} vendors\footnotemark[2] as summarized in
Table~\ref{tab:appDataset}.
Our dataset contains~\totalfiles{} unique files (based on their MD5 hash) as shown in 
Figure~\ref{fig:nb-files-per-vendor} for selected vendors. For each device we plot 
three dots, one for each type of file, while the shape indicates the major Android 
version that the device is running.\footnotemark[3]
The number of pre-installed files varies greatly from one vendor to another. Although 
it is not surprising to see a large amount of native libraries due to hardware differences, 
some vendors embed hundreds of extra apps (\ie ``\texttt{.apk}'' files) compared to other 
manufacturers running the same Android version. 
For the rest of our study, we focus on \totalapps{} Android 
apps present in the dataset\@,
leaving the analysis of root certificates and libraries for future work.

\footnotetext[1]{We use the MD5 hash of the IMEI to uniquely identify a user, and the
build fingerprint reported by the vendor 
to uniquely identify a given device model. Note that two devices with the
same fingerprint may be customized and therefore, have different apps pre-installed.}
\footnotetext[2]{We rely on the vendor string self-reported by the OS vendor, which
could be bogus. For instance, Alps rebrands as ``iPhone'' some of its models, which,
according to information available online, are Android-based replicas of iOS.}
\footnotetext[3]{We found that \noActNoSrvNoRcv{} of the apps do not have any
activity, service, or receiver. These apps may potentially be used as providers
of resources (\eg{} images, fonts) for other apps.}
\footnotetext[4]{We consider that a given device is rooted according to three
  signals. First, when \appname{} has finished the upload of pre-installed
  binaries, 
  the app asks the user whether the handset is rooted according to their own understanding 
  (note that the user may choose not to answer the question). As a complement, we use the 
  library RootBeer~\cite{rootbeer} to progammatically check if a device is rooted or not. 
  If any of these sources indicates that the device is potentially rooted, we consider 
  it as such. Finally, we discard devices where there is evidence of 
  custom ROMs having been installed (\eg LineageOS). We discuss the limitations
of this method in Section~\ref{sec:limitations}.}

Our user-base is geographically distributed across \locationsall{} countries, yet 
\usersEurope{} of our users are located in Europe, \usersAmerica{} in America (North 
and South), and \usersAsia{} in Asia. 
Further, up to \samsungmodelsperc{} and \huaweimodelsperc{} of the total number of 
devices in our dataset belong to Samsung and Huawei ones, respectively. 
This is coherent with market statistics available online~\cite{androidmarketshare0,androidmarketshare1}.  
While both  
manufacturers are Google-certified vendors, our dataset also contains low-end 
Android devices from manufacturers 
targeting markets such as Thailand, Indonesia, and India -- many of these
vendors are not Google-certified. 
Finally, to avoid introducing any bias in our results, 
we exclude \rootedDevices{} potentially rooted handsets from our
study.\footnotemark[4]

\setcounter{footnote}{4}

\begin{figure*}[t]
  \centering
  \footnotesize
  \resizebox{\linewidth}{!}{%
    \begin{tabular}{lllrrrrrrrrr}
      \toprule
	    \multicolumn{1}{c}{\textbf{Vendor}} &
	    \multicolumn{1}{c}{\textbf{Country}} &
	    \multicolumn{1}{c}{\specialcell{\textbf{Certified}\\\textbf{partner}}} &
	    \multicolumn{1}{c}{\specialcell{\textbf{Device}\\\textbf{Fingerprints}}} &
	    \multicolumn{1}{c}{\specialcell{\textbf{Users}}} &
	    \multicolumn{1}{c}{\specialcell{\textbf{Files}\\\textbf{(med.)}}} &
	    \multicolumn{1}{c}{\specialcell{\textbf{Apps}\\\textbf{(med.)}}} &
	    \multicolumn{1}{c}{\specialcell{\textbf{Libs}\\\textbf{(med.)}}} &
	    \multicolumn{1}{c}{\specialcell{\textbf{DEX}\\\textbf{(med.)}}} &
	    \multicolumn{1}{c}{\specialcell{\textbf{Root certs}\\\textbf{(med.)}}} &
	    \multicolumn{1}{c}{\specialcell{\textbf{Files}\\\textbf{(total)}}} &
	    \multicolumn{1}{c}{\specialcell{\textbf{Apps}\\\textbf{(total)}}} \\
      \midrule
      \textbf{Samsung} & South Korea & \samsunggcp{} & \devssamsung{} &
        \userssamsung{} & \filessamsung{} & \appssamsung{} &
        \libssamsung{} & \dexssamsung{} & \certssamsung{} &
        \samsungfiles{} & \samsungapps{} \\
      \textbf{Huawei} & China & \huaweigcp{} & \devshuawei{} &
        \usershuawei{} & \fileshuawei{} & \appshuawei{} &
        \libshuawei{} & \dexshuawei{} & \certshuawei{} &
        \huaweifiles{} & \huaweiapps{} \\
      \textbf{LGE} & South Korea & Yes & \devslge{} &
        \userslge{} & \fileslge{} & \appslge{} &
        \libslge{} & \dexslge{} & \certslge{} &
        \lgefiles{} & \lgeapps{} \\
      \textbf{Alps Mobile} & China & \alpsgcp{} & \devsalps{} &
        \usersalps{} & \filesalps{} & \appsalps{} &
        \libsalps{} & \dexsalps{} & \certsalps{} &
        \alpsfiles{} & \alpsapps{} \\
      \textbf{Motorola} & US/China & Yes & \devsmotorola{} &
        \usersmotorola{} & \filesmotorola{} & \appsmotorola{} &
        \libsmotorola{} & \dexsmotorola{} & \certsmotorola{} &
        \motorolafiles{} & \motorolaapps{} \\
      \midrule
      \textbf{Total (\totalvendors{} vendors)} & \textbf{---} &
        \textbf{\devicesGPtrotect{}} & \textbf{\totaldevices{}} &
        \textbf{\totalusers{}} & & & & & &
        \textbf{\totalfiles{}} & \textbf{\totalapps{}} \\
      \bottomrule
    \end{tabular}
  }\captionof{table}{General statistics for the top-5 vendors in our dataset.}%
\label{tab:appDataset}
\vspace*{-2em}
\end{figure*}

\vspace*{-1em}
\subsection{\lumenname{}}%
\label{sec:lumen}

\lumenname{} is an Android app available on Google Play that aims to promote mobile
transparency and enable user control over their personal data and traffic.
It leverages the Android VPN permission to intercept and analyze all Android traffic 
in user-space and in-situ, even if encrypted, without needing root permissions. By 
running locally on the user's device, \lumenname{} is able to correlate traffic
flows with system-level information and app activity. \lumenname{}'s architecture 
is publicly available and described in~\cite{lumen}. \lumenname{} allows us to 
accurately determine which app is responsible for an observed PII leak from the vantage 
point of the user and as triggered by real user and device stimuli in the wild. 
Since all the analysis occurs on the device, only processed 
traffic metadata is exfiltrated 
from the device. 

\noindent \textbf{Dataset:} 
For this study, we use anonymized traffic logs provided by over \totalLumenUsers{} 
users from \totalLumenCountries{} countries (according to Google Play Store 
statistics) coming from Android phones manufactured by \totalLumenVendors{} vendors.
This includes \totalLumenFlows{} traffic flows from \totalLumenApps{} unique 
apps (\totalLumenAppVersions{} unique package name and version combinations).
However, as \lumenname{} does not collect app fingerprints or hashes of files,
to find the overlap between the \lumenname{} dataset and the pre-installed
apps, we match records sharing the same package name, app version, 
and device vendor as the ones in the pre-installed apps dataset. While this 
method does not guarantee that the overlapping apps are exactly the same,
it is safe to assume that phones that are not rooted are not shipped with
different apps under the same package names and app versions. As a result, we 
have \appsInLumenFlows{} unique pre-installed app/version/vendor combinations 
present in both datasets.

\vspace*{-1em}
\subsection{Ethical Concerns}%
\label{sub:ethics}

Our study involves the collection of data from real users who organically installed
\appname{} or \lumenname{} on their devices. Therefore, we follow the principles of 
informed consent~\cite{menloreport} and we avoid the collection of any personal or sensitive 
data. We sought the approval of our institutional Ethics Board and Data Protection 
Officer (DPO) before starting the data collection. Both tools also provide
extensive privacy policies in their Google Play profile. 
Below we discuss details specific to each tool.

\noindent \textbf{\appname:} 
The app collects some metadata about the device to attribute observations to manufacturers 
(\eg{} its model and fingerprint) along with some data about the 
pre-installed applications (extracted from the Package Manager), network
operator (MNO), and user (the timezone, and the MCC and MNC codes from their SIM card, if available). 
We compute the MD5 hash of the device's IMEI to identify duplicates and updated firmware 
versions for a given device. 

\noindent \textbf{\lumenname:} 
Users are required to opt in twice before initiating traffic interception~\cite{menloreport}. 
\lumenname{} preserves its users' privacy by performing flow processing and analysis on the
device, only sending anonymized flow metadata for research purposes. \lumenname{}
does not send back any unique identifiers, device fingerprints, or raw traffic captures. To further 
protect user's privacy, \lumenname{} also ignores all flows generated by browser
apps  
which may potentially deanonymize a user; and allows the user to disable traffic interception 
at any time.

\vspace*{-1em}
\section{Ecosystem Overview}%
\label{sec:ecosystem}
The openness of Android OS has enabled a complex supply chain 
ecosystem formed by different stakeholders, 
be it manufacturers, MNOs, affiliated developers, and distributors. These actors can add 
proprietary apps and features to Android devices, seeking to provide a better
user experience, add value to their products,  
or provide access to proprietary services. 
However, this could also be for (mutual) financial gain~\cite{facebook,newley}. 
This section provides an overview of pre-installed Android packages to uncover some of 
the gray areas that surround them, the large and diverse set of developers involved, 
the presence of third-party advertising and tracking libraries, 
and the role of each stakeholder.

\vspace*{-1em}
\subsection{Developer Ecosystem}
\label{sec:develecosystem}

\begin{table*}[t]
  \footnotesize
  \centering
  \resizebox{\textwidth}{!}{
    \begin{tabular}{lrll}
      \toprule
	    \multicolumn{1}{c}{\textbf{Company name}} &
	    \multicolumn{1}{c}{\specialcell{\textbf{Number of}\\\textbf{certificates}}} &
	    \multicolumn{1}{c}{\textbf{Country}} &
	    \multicolumn{1}{c}{\specialcell{\textbf{Certified}\\\textbf{partner?}}} \\
      \midrule
      Google & 92 & United States & N/A \\ 
      Motorola & 65 & US/China & Yes \\ 
      Asus & 60 & Taiwan & Yes \\ 
      Samsung & 38 & South Korea & Yes \\ 
      Huawei & 29 & China & Yes \\ 
      \midrule
      \textbf{Total (vendors)} & \textbf{\manufacturerCerts{}} & \textbf{---} & \textbf{---} \\       \bottomrule
    \end{tabular}
    \hspace{\columnsep}
    \begin{tabular}{lrlr}
      \toprule
	    \multicolumn{1}{c}{\textbf{Company name}} &
	    \multicolumn{1}{c}{\specialcell{\textbf{Number of}\\\textbf{certificates}}} &
	    \multicolumn{1}{c}{\textbf{Country}} &
	    \multicolumn{1}{c}{\specialcell{\textbf{Number of}\\\textbf{vendors}}} \\
      \midrule
      MediaTek & 19 & China & \nbVendorsMediatek{} \\
      Aeon & 12 & China & \nbVendorsAeon{} \\
      Tinno Mobile & 11 & China & \nbVendorsTinno{} \\
      Verizon Wireless & 10 & United States & \nbVendorsVerizon{} \\
      \textit{Unknown company} & 7 & China & \nbVendorsUnknown{} \\
      \midrule
      \textbf{Total} & \textbf{\otherCerts{}} & \textbf{---} & \textbf{\totalvendors{}}\\
      \bottomrule
    \end{tabular}
  }
	\caption{\textbf{Left:} top-5 most frequent developers (as per the
	total number of apps signed by them),
  and \textbf{right:} for other companies.}
\label{tab:apps-devs}
\vspace*{-1em}%
\end{table*}

We start our study by analyzing the organizations signing each pre-installed app. First, we cluster 
apps by the unique certificates used to sign them and then we rely on the information present in 
the \texttt{Issuer} field of the certificate to identify the organization~\cite{signedapps}. Despite 
the fact that this is the most reliable signal to identify the organization signing the software, it 
is still noisy as a company can use multiple certificates, one for each organizational unit. More importantly, these are self-signed certificates, which significantly lowers the trust that can be put on them.

We were unable to identify the company behind several certificates (denoted as \textit{Unknown company} 
in Table~\ref{tab:apps-devs}) due to insufficient or dubious information in the certificate: \eg{} the 
\texttt{Issuer} field only contains the mentions \texttt{Company} and \texttt{department}. We have come 
across apps that are signed by 42 different \textit{"Android Debug"} certificates on phones from 21 
different brands. This reflects poor and potentially insecure development practices as Android's debug 
certificate is used to automatically sign apps in development environments, hence enabling other apps 
signed with that certificate to access its functionality without requesting any permission. 
Most app 
stores (including Google Play) will not accept the publication of an app signed with 
a Debug certificate~\cite{androidDev}.
Furthermore, we also found as many as~\androidCerts{} certificates that only mention \textit{``Android''} 
in the \texttt{Issuer} field. A large part (\androidCertsUS{}) of those certificates are supposedly issued 
in the US, while others seem to have been issued in Taiwan (\androidCertsTW{}), China (\androidCertsCN{}),
and Switzerland (\androidCertsCH{}). In the absence of a public list of official developer certificates,
it is not possible to verify their authenticity or know their owner, as discussed in Section~\ref{sec:limitations}.

With this in mind, we extracted \totalCerts{} unique certificates out of our dataset. Table~\ref{tab:apps-devs} 
shows the 5 most present companies in the case of phone vendors (left) and other development companies (right).
This analysis uncovered a vast landscape of third-party software in the long-tail, including large digital 
companies (\eg LinkedIn, Spotify, 
and TripAdvisor), as well as advertising and tracking services. This is the case of ironSource, an advertising
firm signing pre-installed software~\cite{ironsrc} found in Asus, Wiko and other vendors, 
and TrueCaller, a service to block unwanted call or texts~\cite{truecaller}. 
\claim{According to their website and also independent
sources~\cite{truecallerdata1, truecallerdata2},
TrueCaller uses crowdsourced mechanisms to build
a large dataset of phone numbers used for spam and also for advertising}. 
Likewise, we have found 123 apps (by their MD5) signed by Facebook. 
These apps are found in 939 devices, \fbAppsOnSamsung{} of which are Samsung's.
We have also found apps signed by AccuWeather, \claim{a weather service
previously found
collecting personal data aggressively~\cite{ren2018bug}},
Adups software, responsible for the Adups backdoor~\cite{kryptowireADUPS}, and GMobi~\cite{gmobi}, 
a mobile-advertising 
company previously accused of dubious practices by the Wall Street Journal~\cite{newley}.

\vspace*{-1em}
\subsection{Third-party Services}
As in the web, mobile app developers can embed in their pre-installed software 
third-party libraries (TPLs) provided 
by other companies, including libraries (SDKs) provided by 
ad networks, analytics services or social networks. In
this section we use LibRadar++, an obfuscation-resilient tool to identify
TPLs used in Android apps~\cite{libradarplusplus}, on our dataset to
examine their presence due to the potential privacy implications for users:
when present in pre-installed apps, TPLs have the capacity to 
monitor user's activities longitudinally~\cite{vallina2012breaking,razaghpanah2018apps}.
We exclude well-known TPLs providing development support such as the Android support
library. 
First, we classify the~\libsAll{} unique TPLs identified by LibRadar++ according to the 
categories reported by Li~\etal{}~\cite{li2016investigation}, AppBrain~\cite{appbrain}, 
and PrivacyGrade~\cite{privacygrade}. 
We manually classified those TPLs that were not categorized by these datasets.

\begin{figure}[t]
  \centering
  \footnotesize
  \scalebox{1.0}{%
    \begin{tabular}{lrrrl}
      \toprule
	    \multicolumn{1}{c}{\textbf{Category}} &
	    \multicolumn{1}{c}{\specialcell{\textbf{\# libraries}}} &
	    \multicolumn{1}{c}{\specialcell{\textbf{\# apps}}} & 
	    \multicolumn{1}{c}{\specialcell{\textbf{\# vendors}}} &
	    \multicolumn{1}{c}{\specialcell{\textbf{Example}}} \\
      \midrule
	    Advertisement & \libAdvertisementNbLibs{}
      (\libAdvertisementNbLibsGrouped{}) & \libAdvertisementNbApps{} & 164 & Braze \\
	    Mobile analytics & \libAnalyticsNbLibs{} (\libAnalyticsNbLibsGrouped{}) & \libAnalyticsNbApps{} & 158 & Apptentive \\
	    Social networks & \libSocialNbLibs{} (\libSocialNbLibsGrouped{}) & \libSocialNbApps{} & 157 & Twitter \\
      \midrule
	    \textbf{All categories} & \textbf{\libsIded{}} & \textbf{\appsWithLibs{}} & \textbf{165} &  \textbf{---} \\
      \bottomrule
    \end{tabular}
  }\captionof{table}{%
  Selected TPL categories present in pre-installed apps. In 
    brackets, we report the number of TPLs when grouped by package name. 
  }%
\label{tab:third-party-libs}
\vspace*{-2em}%
\end{figure}

We focus on categories that could cause harm to the users' privacy, such as mobile analytics 
and targeted advertisement libraries.  
We find \libsIded{} TPLs in such categories, as summarized in
Table~\ref{tab:third-party-libs}. We could identify \claim{advertising and tracking} 
companies such as Smaato (specialized 
in geo-targeted ads~\cite{smaatoGeo}),
GMobi, Appnext, ironSource, Crashlytics, and Flurry. Some of these third-party providers were
also found shipping their own packages in Section~\ref{sec:develecosystem} or are prominent 
actors across apps published in Google Play Store~\cite{razaghpanah2018apps}.
We found~\appsWithFacebook{} apps embedding Facebook's 
Graph SDK which is distributed over \fgsWithFacebook{} devices. 
The certificates of these apps suggests that~\vendorFBApps{} of them were signed by the device 
vendor, and~\ispFBApps{} by an operator (only \FBApps{} are signed by Facebook itself).
The presence of Facebook's SDKs in pre-installed apps could, in some cases,
\claim{be explained by partnerships established by Facebook with Android
vendors as the New York Times revealed~\cite{facebook}}.

We found other companies that provide mobile analytics and app monetization schemes such as Umeng, 
Fyber (previously Heyzap), and Kochava~\cite{razaghpanah2018apps}. 
Moreover, we also found instances of advanced analytics 
companies in Asus handsets such as Appsee~\cite{appsee} and Estimote~\cite{estimote}. 
\claim{According to their website, 
Appsee is a TPL that allows developers to record and upload the users'
screen~\cite{appseerecords}, including touch 
events~\cite{pan2018panoptispy}. If, by itself, recording the user's screen does not constitute 
a privacy leak, recording and uploading this data could unintentionally 
leak private information such as account details}. Estimote develops solutions for indoors geo-localization~\cite{estimote}.
Estimote's SDK allows an app to react to nearby wireless beacons to, for example, send personalized push 
notifications to the user upon entering a shop

Finally, we find TPLs provided by companies 
specialized in the Chinese market~\cite{libradarplusplus} in 548 pre-installed 
apps.
The most relevant ones are Tencent's SDK, AliPay (a payment service) and Baidu SDK~\cite{baidusdk} 
(for advertising and geolocation / geo-coding services), \claim{the last two
possibly used as replacements
for Google Pay and Maps in the Chinese market, respectively}. Only one of the apps embedding these 
SDKs is signed by the actual third-party service provider, which indicates that their presence in pre-installed 
apps is likely due to the app developers' design decisions.

\vspace*{-1em}
\subsection{Public and Non-public Apps}

We crawled the Google Play Store to identify how many of the pre-installed apps found by \appname{} are available to
the public. This analysis took place on the 19th of November, 2018 and we only used the package name of the pre-installed apps 
as a parameter.
We found that only \pkgOnGPlay{} of the package names in our dataset are indexed in the Google Play Store. 
For those indexed, few categories dominate the spectrum of pre-installed
apps according to Google Play metadata, notably communication, 
entertainment, productivity, tools, and multimedia apps.

The low presence of pre-installed apps in the store suggests that this type of software might have escaped any scrutiny 
by the research community.
In fact, we have found samples of pre-installed apps developed by prominent organizations that are not publicly available 
on Google Play. For instance, software developed and signed by Facebook (\eg{} \texttt{com.facebook.appmanager}), Amazon, and 
CleanMaster among others. Likewise, we found non-publicly available versions of popular web browsers (\eg{} UME Browser, Opera).

Looking at the last update information reported by Android's package manager for
these apps, 
we found that pre-installed apps also present on 
Google Play are updated more often than the rest of pre-installed apps:
\appsNeverUpdatedPerc{} of the non-public apps do not seem to get updated and 
\appsNeverUpInFiveYearsOrMore{} of them remained unpatched for 5 years or more. 
If a vulnerability exists in one of these 
applications (see Section~\ref{sec:appAnalysis}), the user may stay at risk for 
as long as they keep using the device.

\vspace*{-1em}
\section{Permission Analysis}%
\label{sec:permissions}

Android implements a permissions model to control apps' access to sensitive data 
and system resources~\cite{perms}. 
By default, apps are not allowed to 
perform any protected operation. Android permissions are not limited to those defined by AOSP: 
any app developer -- including manufacturers -- can define their own \textit{custom permissions} 
to expose their functionality to other apps~\cite{customperms}. We leverage Androguard~\cite{androguard} 
to extract and study the permissions, both declared and requested, by pre-installed apps. We 
primarily focus on custom permissions as $i)$ pre-installed services have privileged access to 
system resources, and $ii)$ \claim{privileged pre-installed services may (involuntarily) expose critical 
services and data, even bypassing Android's official permission set.}

\vspace*{-.5em}
\subsection{Declared Custom Permissions}
\label{sec:custompermissions}

We identify \totalPackagesDeclarigCustomPerms{} unique Android package names across 
\totalVendorsDeclarigCustomPerms{} Android vendors defining \totalCustomPerms{} custom 
permissions. We exclude AOSP--defined permissions and those associated with Google's 
Cloud Messaging (GCM)~\cite{gcm_perms}.
The number of custom permissions declared per Android vendor varies across brands and 
models due to the actions of other stakeholders in the supply chain. 
We classify the organizations declaring 
custom permissions in 8 groups as shown in Table~\ref{tab:perm-coresystems}: 
hardware vendors, MNOs (\eg{} Verizon), third-party services (\eg{} Facebook), AV firms 
(\eg{} Avast), industry alliances (\eg{} GSMA), chipset manufacturers (\eg{} Qualcomm), 
and browsers (\eg{} Mozilla). We could not confidently identify the organizations responsible 
for \percentageUnclassifiedPerms{} of all the custom permissions.%
\footnote{While Android's documentation recommends using reverse-domain-style naming for defining  
custom permissions to avoid collisions.~\cite{customperms}, 
\totalNonOfficialPermsWithAOSPrefix{} of them -- many of which are declared by a single hardware 
vendor -- start with AOSP prefixes such as \textit{android.permission.*}. The absence of good 
development practices among developers 
complicated this classification, forcing us to follow a semi-manual process 
that involved analyzing multiple signals to identify their possible 
purpose and for attribution.}

\begin{table*}[t]
  \centering
  \resizebox{\textwidth}{!}{%
	\begin{tabular}{lccccccccc}
      	\toprule
	 & \textbf{Custom} &
	\multicolumn{8}{c}{\textbf{Providers}}
    \\ \cmidrule[.05em]{3-10}
	& \textbf{permissions} & 
	\textbf{Vendor} & 
	\textbf{Third-party} &
	\textbf{MNO} &
	\textbf{Chipset} &
	\textbf{AV / Security} &
	\textbf{Ind. Alliance} &
	\textbf{Browser} &
	\textbf{Other} 
	\\ \midrule
	\textbf{Total} &
    \totalCustomPerms{} (\totalVendorsCustomPerms{}) &
    \totalNonOfficialVendorTotalPerms{} (\totalNonOfficialVendorTotalVendors{}) & 
    \totalNonOfficialtpTotalPerms{} (\totalNonOfficialtpTotalVendors{}) & 
    \totalNonOfficialMNOTotalPerms{} (\totalNonOfficialMNOTotalVendors{}) & 
    \totalNonOfficialchipsetTotalPerms{} (\totalNonOfficialchipsetTotalVendors{}) &
    \totalNonOfficialavSecTotalPerms{} (\totalNonOfficialavSecTotalVendors{}) &
    \totalNonOfficialAllianceTotalPerms{} (\totalNonOfficialAllianceTotalVendors{}) &
    \totalNonOfficialbrowserTotalPerms{} (\totalNonOfficialbrowserTotalVendors{}) &
    \totalNonOfficialOtherTotalPerms{} (\totalNonOfficialOtherTotalVendors{}) 
    \\[.5em]
	\textbf{Android Modules} & & & & 
		\\ \midrule
	    \texttt{android} &  
    \totalAndroidCustomPerms{} (\totalAndroidCustomPermsVendors{}) &
    \totalAndroidVendorCustomPerms{} (\totalAndroidVendorCustomPermsVendors{}) & 
		--- & 
    \totalAndroidMNOCustomPerms{} (\totalAndroidMNOCustomPermsVendors{}) &
    \totalAndroidChipsetCustomPerms{} (\totalAndroidChipsetCustomPermsVendors{}) &
		--- &
    \totalAndroidAllianceCustomPerms{} (\totalAndroidAllianceCustomPermsVendors{}) & 
		--- &
    \totalAndroidOTHERCustomPerms{} (\totalAndroidOTHERCustomPermsVendors{}) \\
	    \texttt{com.android.systemui} & 
    \totalAndroidUICustomPerms{} (\totalAndroidUICustomPermsVendors{}) &
    \totalAndroidUIVendorCustomPerms{} (\totalAndroidUIVendorCustomPermsVendors{}) &  
    \totalAndroidUIThirdPCustomPerms{} (\totalAndroidUIThirdPCustomPermsVendors{}) & 
		--- &
		--- &
		--- &
		--- &
		--- &
    \totalAndroidUIOTHERCustomPerms{} (\totalAndroidUIOTHERCustomPermsVendors{}) \\
	    \texttt{com.android.settings} &
    \totalAndroidSettingsCustomPerms{} (\totalAndroidSettingsCustomPermsVendors{}) &
    \totalAndroidSettingsVendorCustomPerms{} (\totalAndroidSettingsVendorCustomPermsVendors{}) & 
		---  & 
    \totalAndroidSettingsMNOCustomPerms{} (\totalAndroidSettingsMNOCustomPermsVendors{}) &
		--- &
		--- & 
		--- & 
		--- &
    \totalAndroidSettingsOTHERCustomPerms{} (\totalAndroidSettingsOTHERCustomPermsVendors{}) \\
	    \texttt{com.android.phone} &  
    \totalAndroidPhoneCustomPerms{} (\totalAndroidPhoneCustomPermsVendors{}) & 
    \totalAndroidPhoneVendorCustomPerms{} (\totalAndroidPhoneVendorCustomPermsVendors{}) & 
		---  & 
    \totalAndroidPhoneMNOCustomPerms{} (\totalAndroidPhoneMNOCustomPermsVendors{}) &
    \totalAndroidPhoneChipsetCustomPerms{} (\totalAndroidPhoneChipsetCustomPermsVendors{}) &
		--- & 
		--- &
		--- &
    \totalAndroidPhoneOTHERCustomPerms{} (\totalAndroidPhoneOTHERCustomPermsVendors{}) \\
	    \texttt{com.android.mms} &  
    \totalAndroidMMSCustomPerms{} (\totalAndroidMSSCustomPermsVendors{}) & 
    \totalAndroidMMSVendorCustomPerms{} (\totalAndroidMMSVendorCustomPermsVendors{}) & 
		---  & 
    \totalAndroidMMSMNOCustomPerms{} (\totalAndroidMMSMNOCustomPermsVendors{}) &
		--- &
		--- & 
    \totalAndroidMMSAllianceCustomPerms{} (\totalAndroidMMSAllianceCustomPermsVendors{}) &
		--- &
    \totalAndroidMMSOTHERCustomPerms{} (\totalAndroidMMSOTHERCustomPermsVendors{}) \\
	    \texttt{com.android.contacts} &
    \totalAndroidContactsCustomPerms{} (\totalAndroidContactsCustomPermsVendors{}) & 
    \totalAndroidContactsVendorCustomPerms{} (\totalAndroidContactsVendorCustomPermsVendors{}) & 
		--- & 
		--- &
		--- &
		--- &
		--- &
		--- &
    \totalAndroidContactsOTHERCustomPerms{} (\totalAndroidContactsOTHERCustomPermsVendors{}) \\

	   \texttt{com.android.email} & 
    \totalAndroidEmailCustomPerms{} (\totalAndroidEmailCustomPermsVendors{}) & 
    \totalAndroidEmailVendorCustomPerms{} (\totalAndroidEmailVendorCustomPermsVendors{}) & 
		--- &
		--- &
		--- &
		--- &
		--- &
		--- &
    \totalAndroidEmailOTHERCustomPerms{} (\totalAndroidOTHERCustomPermsVendors{}) \\

    \bottomrule
    \end{tabular}}
\caption{\label{tab:perm-coresystems}
	Summary of custom permissions per provider category and their presence
	in selected sensitive Android core modules. 
	The value in brackets reports the number of Android vendors in which
	custom permissions were found.}
\vspace*{-1.5em}%
\end{table*}

As shown in Table~\ref{tab:perm-coresystems}, \percentageVendorTotalPerms{} of all declared
custom permissions are defined by \totalNonOfficialVendorTotalProviders{} handset vendors
according to our classification. 
\claim{Most of them are associated with proprietary services such as Mobile Device Management (MDM) 
solutions for enterprise customers.}
Yet three vendors account for over 68\% of the total custom permissions; namely Samsung 
(\percentageSamsungVendorTotalPerms{}), Huawei (\percentageHuaweiVendorTotalPerms{}), and
Sony (formerly Sony-Ericsson, \percentageSonyVendorTotalPerms{}).
Most of the custom permissions added by hardware vendors
-- along with chipset manufacturers, and MNOs -- are exposed by Android core services, including 
the default browser \texttt{com.android.browser}. Unfortunately, as demonstrated in the MediaTek 
case~\cite{felt2011permission}, exposing such sensitive resources in critical services may potentially increase the attack 
surface if not implemented carefully.

\claim{An exhaustive analysis of custom permissions also suggests
(and in some cases confirms) the presence of service 
integration and commercial partnerships between handset vendors, MNOs, analytics 
services (\eg{}  Baidu, ironSource, and Digital Turbine), and online 
services (\eg{} Skype, LinkedIn, Spotify, CleanMaster, and Dropbox).} 
We also found custom permissions associated with vulnerable modules (\eg{} MediaTek)
and potentially \claim{harmful} services (\eg{} Adups). 
We discuss cases of interest below.

\noindent \textbf{VPN solutions:} 
Android provides native support to third-party VPN clients. This feature is considered 
as highly sensitive as it gives any app requesting access the capacity to break 
Android's sandboxing and monitor users' traffic~\cite{AndroidVPNService, ikram2016analysis}. 
The analysis of custom permissions reveals that Samsung and Meizu implement their own VPN 
service. It is unclear why these proprietary VPN implementations exist but it has been 
reported as problematic by VPN developers for whom their clients, designed for Android's
default VPN service, do not run on such handsets~\cite{adguardMeizu,lumen,ikram2016analysis}.
A complete analysis of these VPN packages is left for future work.

\noindent \textbf{Facebook:}
We found 6 different Facebook packages, three of them unavailable on Google Play, declaring 
\totalFBPerms{} custom permissions as shown in Table~\ref{tab:fbPerms}. 
These permissions have been found in \totalFBVendors{} Android vendors, including Samsung, 
Asus,
Xiaomi, HTC, Sony, and LG.
According to users' complaints, two of these packages (\texttt{com.facebook.appmanager} and 
\texttt{com.facebook.system}) seem to automatically download other Facebook software such 
as Instagram in users' phones~\cite{androidCentralFacebookAppManager,XDAFacebookSystem}.
\claim{We also found interactions between Facebook and MNOs such as Sprint.}

\begin{table}[t]
  \centering
  \resizebox{\linewidth}{!}{%
    \begin{tabular}{l ccc }
      \toprule
      \textbf{Package} & \textbf{Public} &
      \textbf{\# Vendors} &
      \textbf{\# Permissions} \\
      \midrule
      \texttt{com.facebook.system} & No & \customcomfacebooksystemVendors{} & \customcomfacebooksystemPermissions{} \\
      \texttt{com.facebook.appmanager} & No & \customcomfacebookappmanagerVendors{} & \customcomfacebookappmanagerPermissions{} \\
      \texttt{com.facebook.katana} (Facebook) & Yes & \customcomfacebookkatanaVendors{} & \customcomfacebookkatanaPermissions{} \\
      \texttt{com.facebook.orca} (Messenger) & Yes & \customcomfacebookorcaVendors{} & \customcomfacebookorcaPermissions{} \\
      \texttt{com.facebook.lite} (FB Lite) & Yes & \customcomfacebookliteVendors{} & \customcomfacebooklitePermissions{} \\
      \texttt{com.facebook.pages.app} & No & \customcomfacebookpagesappVendors{} & \customcomfacebookpagesappPermissions{} \\
      \midrule
      \textbf{Total} & 3 & \totalFBVendors  & \totalFBPerms \\ 
      \bottomrule
    \end{tabular}}
  \caption{\label{tab:fbPerms}
  Facebook packages on pre-installed handsets.}
\vspace*{-2em}%
\end{table}

\noindent \textbf{Baidu:} Baidu's geo-location permission is exposed by
pre-installed apps, including core Android modules, in 
\totalBaiduVendorsCustomPerms{} different vendors, mainly Chinese ones. 
This permission seems to be associated with Baidu's geocoding API~\cite{baidugeocoding} 
and could allow app developers to circumvent Android's location permission.

\noindent \textbf{Digital Turbine:} 
We have identified \totalDTPermissionsCustomPerms{} custom permissions in \totalDTPermissionsCustomPerms{} 
vendors associated with Digital Turbine and its subsidiary LogiaGroup. 
\claim{Their privacy policy indicates that they collect personal data ranging from UIDs to traffic logs that could be
shared with their business partners, which are
undisclosed~\cite{digitalTurbinePolicy}. }
According to the SIM information of these devices, 
Digital Turbine modules are mainly found in 
North-American and Asian users. 
One package name, \texttt{com.dti.att} 
(``dti'' stands for Digital Turbine Ignite), suggests the presence of a partnership with AT\&T. A manual analysis
confirms that this is the case. 
\claim{By inspecting their source-code, this package seems to 
implement comprehensive software management service. Installations 
and removals of apps by users are tracked and linked with PII, which only seem to be ``masked'' (\ie{} hashed)
discretionally.}

\noindent \textbf{ironSource: }
The advertising company ironSource exposes custom permissions related to its AURA Enterprise Solutions~\cite{ironSourceAura}.
We have identified several vendor-specific packages 
exposing custom ironSource permissions, in devices made by vendors such as Asus, Wiko, and HTC (the package name and certificate
signatures suggest that those modules are possibly introduced with vendor's collaboration). 
According to ironSource's material~\cite{ironSourceAuraSlide}, 
AURA has access to over 800 million users per month, 
while gaining access to advanced analytics services and to pre-load software on customers' devices. 
A superficial analysis of some of these packages (\eg{} \texttt{com.ironsource.appcloud.oobe.htc}, \texttt{com.ironsource.appcloud.oobe.asus}) 
reveals that they provide vendor-specific out-of-the-box-experience apps (OOBE) 
to customize a given user's device when the
users open their device for the first time and empower user
engagement~\cite{ironSourceAura}, while also monitoring users' activities.

\noindent \textbf{Other Advertising and Tracking Services:}
Discussing every custom permission introduced by third-party services individually would require an analysis 
beyond the scope of this paper. However, there are a couple of anecdotes of interest that we discuss next. 
One is the case of a pre-installed app signed by Vodafone (Greece) and present in a Samsung device that exposes 
a custom permission associated with Exus~\cite{exusUK}, 
a firm specialized in credit risk management and banking solutions. 
Another service declaring custom permissions in Samsung and LG handsets (likely sold by Verizon) is the analytics 
and user engagement company Synchronoss. 
\claim{Its privacy policy acknowledges the collection, processing and sharing 
of personal data~\cite{synchronossprivacypolicy}. }

\noindent \textbf{Call protection services:}
We identify three external companies providing services for blocking undesired and spam phone calls and text
messages: Hiya~\cite{hiya}, TrueCaller~\cite{truecaller}, and PrivacyStar~\cite{privacystar}. 
Hiya's solution 
seems to be integrated by T-Mobile (US), Orange (Spain), and AT\&T (US) in
\claim{their subsidized} Samsung and LG handsets
according to the package signatures. 
Hiya and TrueCaller's privacy policies \claim{indicate that they collect personal 
data from end users, including contacts stored 
in the device, UIDs, and personal information~\cite{hiyaDataPolicy}}. 
\footnote{Note: the information rendered in their privacy policy differs when crawled from a machine in the EU
or the US\@. As of January 2019, 
none of these companies mention the new European GDPR directive in their privacy policies.} 
PrivacyStar's privacy policy, instead, claims that any information collected 
from a given user's contacts is ``NOT exported 
outside the App for any purpose''~\cite{privacystarpolicy}.

\subsection{Used Permissions}

\begin{figure}[t]
  \centering
  \includegraphics[width=0.9\linewidth]{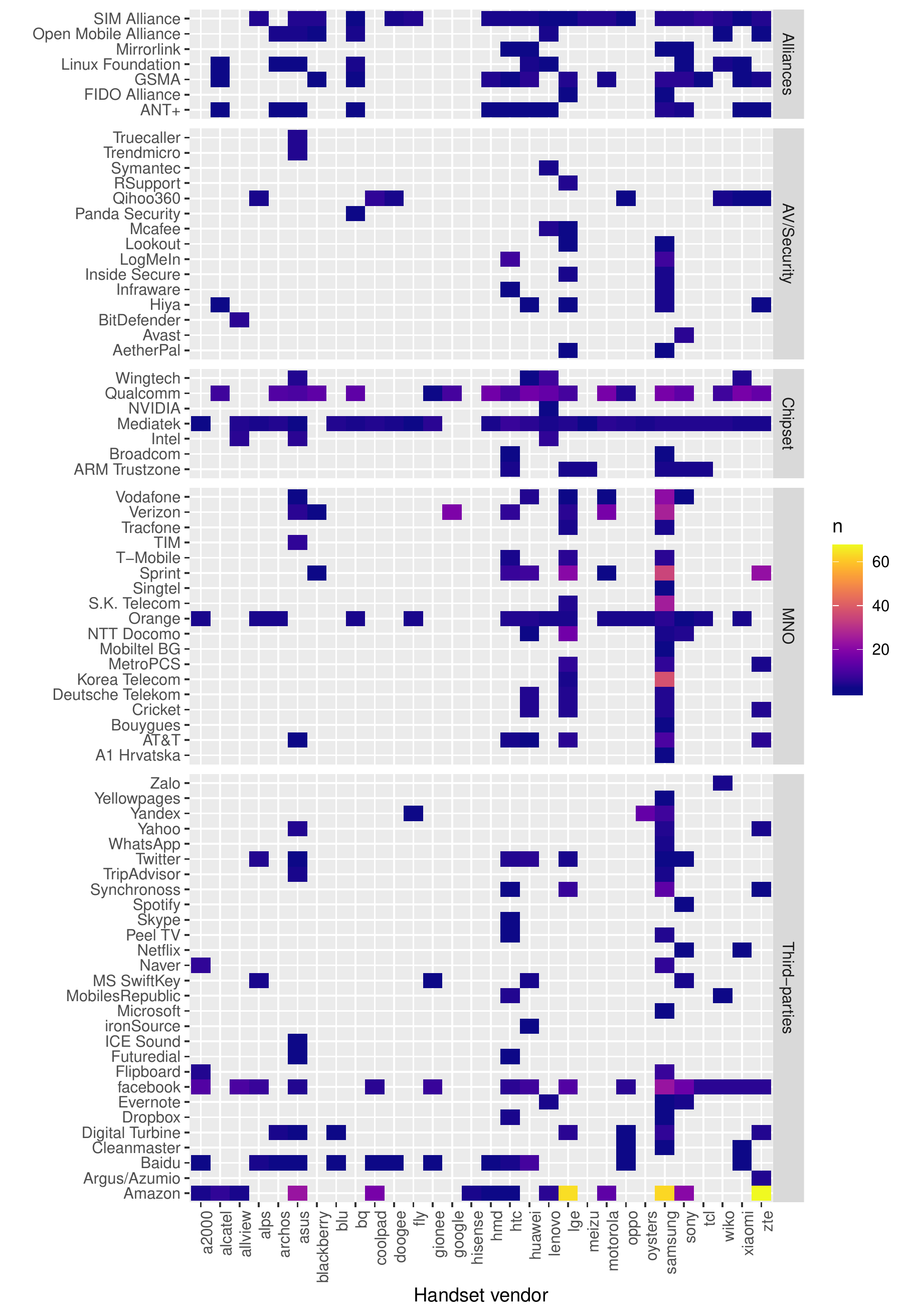}
  \caption{Permissions defined by AV firms, MNOs, 
	chipset vendors and third parties, 
  requested by pre-installed apps.}%
\label{fig:requeseted-permission-vendor}
\vspace*{-2em}%
\end{figure}

\begin{figure}[t]
  \centering
  \includegraphics[width=0.9\linewidth]{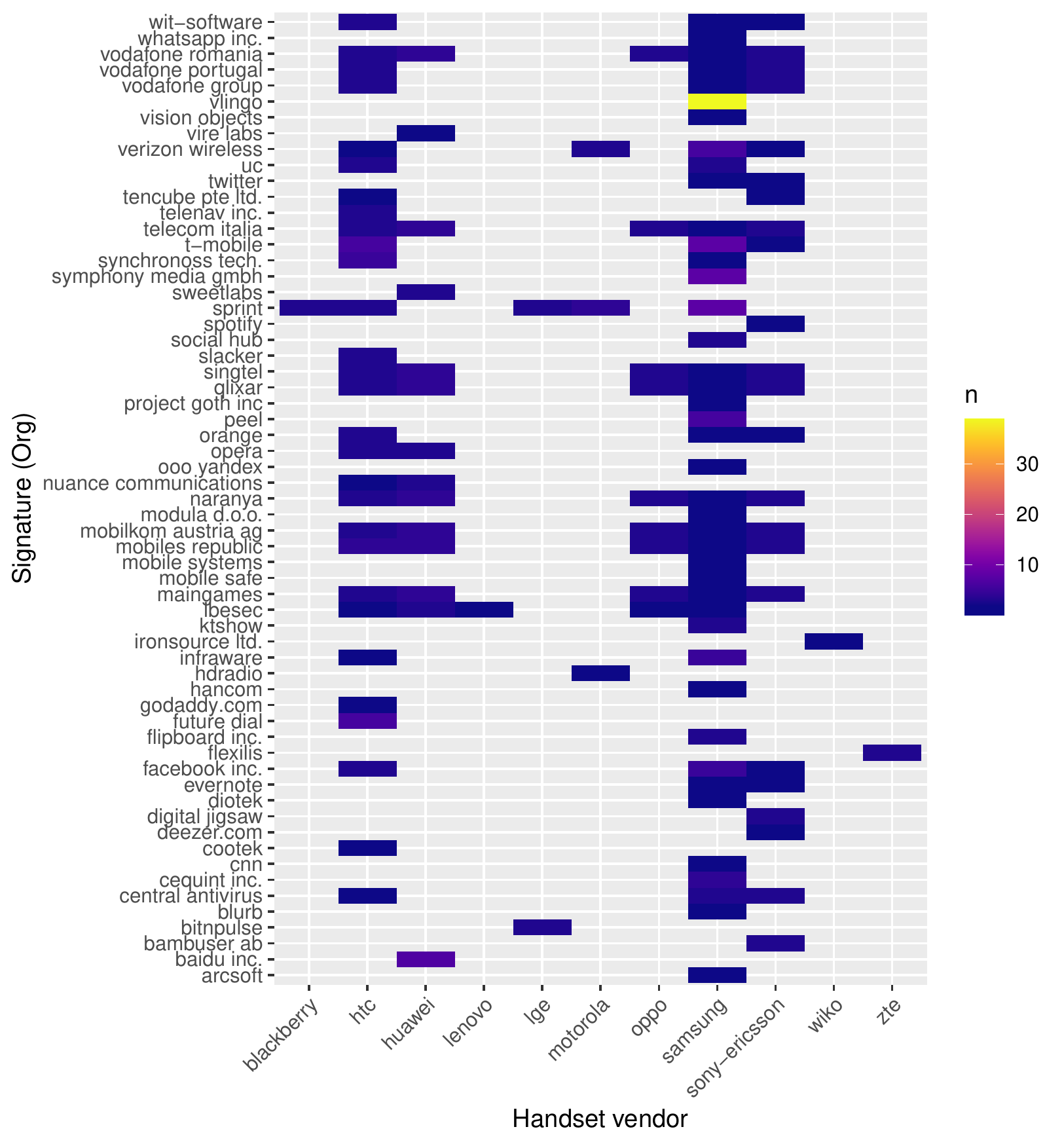}
  \caption{Apps accessing vendors' custom permissions.}%
\label{fig:requeseted-permission-vendor_thirdparty}
\vspace*{-1.5em}%
\end{figure}

The use of permissions by pre-installed Android apps follows a power-law distribution:
\totalAppsUsingPerms{} of the package names request at least one permission and
\totalAppsUsingHundredPerms{} apps request more that 100. The fact that 
pre-installed apps request many permissions to deliver their service does not necessarily 
imply a breach of privacy for the user. 
However, we identified a significant number of \claim{potentially} 
over-privileged vendor- and MNO-specific 
packages \claim{with suspicious activities} such as
\texttt{com.jrdcom.Elabel} -- a package signed by TCLMobile requesting 145 permissions and 
labeled as \claim{malicious} by Hybrid Analysis (a free online malware analysis
service) --and \texttt{com.cube26.coolstore} (144 permissions). 
Likewise, the calculator app found on a Xiaomi Mi 4c requests user's location and the phone 
state, which gives it access to UIDs such as the IMEI\@. We discuss more  
instances of over-privileged apps in Section~\ref{sec:appAnalysis:manual}.

\noindent \textbf{Dangerous Android permissions.}
The median pre-installed Android app requests three dangerous AOSP permissions. 
When we look at the set of permissions requested by a given app (by its package name) across vendors, we 
can notice significant differences. 
We investigate such variations in a subset of 150 package names present at least in 20 different vendors. 
This list contains mainly core Android services as well as apps signed by independent companies (\eg{} Adups) 
and chipset manufacturers (\eg{} Qualcomm). 

Then, we group together all the permissions requested by a given package 
name across all device models for each brand. As in the case of exposed custom permissions, we can see a tendency 
towards over-privileging these modules in specific vendors. For instance, the number of permissions requested 
by the core \texttt{android} module can range from 9 permissions in a Google-branded 
Android device to over 100 in most 
Samsung devices. Likewise, while the median \texttt{com.android.contacts}
service requests 35 permissions, 
this number goes over 100 for Samsung, Huawei, Advan, and LG devices.

\noindent \textbf{Custom permissions.}
2,910 pre-installed apps request at least one custom permission. 
The heatmap in Figure~\ref{fig:requeseted-permission-vendor} 
shows the number of custom permissions requested by pre-installed packages in a hand-picked set of popular Android manufacturers (x-axis).
As we can see, the use of custom permissions also varies across vendors, with those associated with large third-party
analytics and tracking services (\eg{} Facebook), MNOs (\eg{} Vodafone), and AV/Security services (\eg{} Hiya) being
the most requested ones.

This analysis uncovers possible 
partnerships beyond those revealed in the previous sections. We identify vendor-signed services 
accessing ironSource's, Hiya's, and AccuWeather's permissions. 
\claim{This state of affairs potentially allows third-party services and
developers to
gain access to protected permissions requested by other pre-installed packages
signed with the same signature}.
Further, we found 
Sprint-signed packages resembling that of Facebook and Facebook's Messenger APKs (\texttt{com.facebook.orca.vpl} and 
\texttt{com.facebook.katana.vpl}) requesting Flurry-related permissions (a 
third-party tracking service owned by Verizon).

Commercial relationships between third-party services and vendors \claim{appear
to be} bi-directional as shown in Figure~\ref{fig:requeseted-permission-vendor_thirdparty}. 
This figure shows \claim{evidence} of 87 apps accessing vendor permissions,
including packages signed by Facebook, ironSource, 
Hiya,  Digital Turbine, Amazon, Verizon, Spotify, various browser, and MNOs -- 
grouped 
by developer signature for clarity purposes. As the heatmap indicates, Samsung, HTC and Sony are the vendors enabling most of 
the custom permissions requested by over-the-top apps. We found instances
of apps listed on the Play Store also requesting 
such permissions. Unfortunately, custom permissions are not shown to users 
when shopping for mobile apps in the store -- therefore they are apparently 
requested without consent -- allowing 
them to cause serious damage to users' privacy when misused by apps.

\vspace*{-.5em}
\subsection{Permission Usage by TPLs}%
\label{subsec:permissionsLibraries}

\begin{figure}[t]
  \centering
  \includegraphics[width=0.95\linewidth]{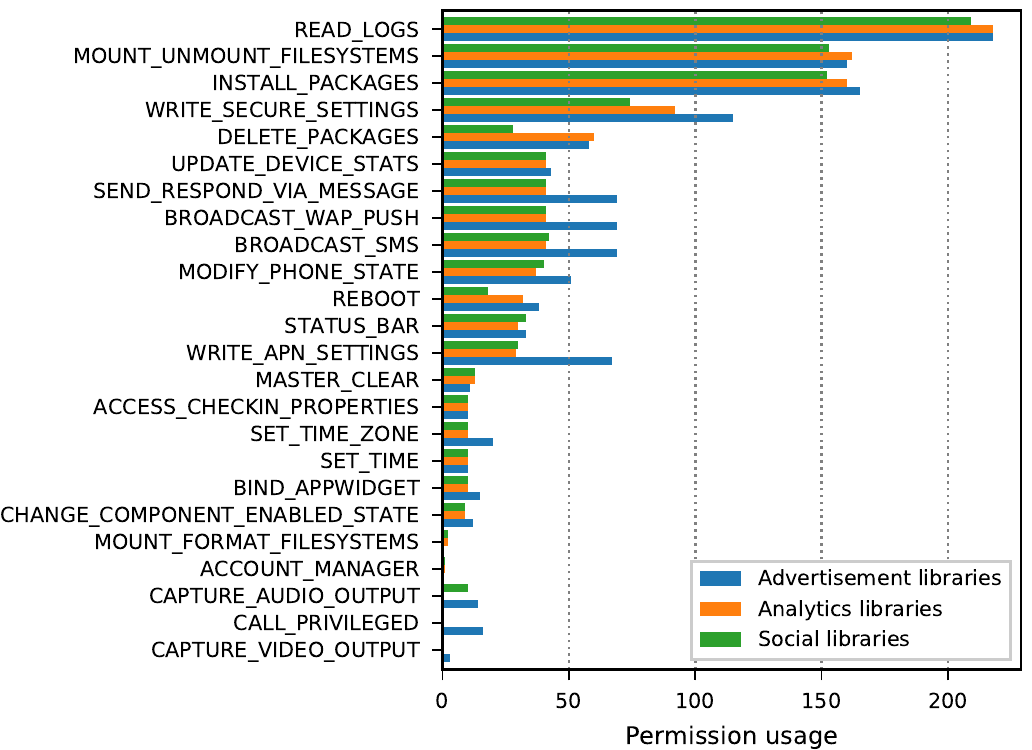}
  \caption{System permissions requested by pre-installed 
	apps embedding TPLs.}%
  \label{fig:third-party-libs-perms}
\vspace*{-1.5em}%
\end{figure}

We look at the permissions used by apps embedding at least one TPL\@. We study the access to permissions with a protection level of either
\texttt{signature} or \texttt{signature|privileged} \claim{as they can only be granted to system apps~\cite{systemPerms} or those signed with a 
system signature}. 
The presence of TPLs in pre-installed apps requesting access to a signature or dangerous permission can, therefore,
give it access to very sensitive resources without user awareness and consent.
Figure~\ref{fig:third-party-libs-perms} shows the distribution of signature
permissions requested across apps embedding TPLs. We find that 
the most used permissions -- \texttt{READ\_LOGS} -- 
allow the app (and thus the TPLs within it) to read system logs, mount and unmount filesystems, or install
packages. We find no significant differences between the three types of TPLs of interest. For completeness, we also find that \appsLibsCustomPerms{} 
apps embedding TPLs of interest request custom permissions as well. Interestingly, 
\libsSamsungCustomPerms{}  of the \libsCustomPerms{} custom permissions 
used by these apps are defined by Samsung.

\vspace*{-.5em}
\subsection{Component Exposing}%
\label{sec:apiexposing}

Custom permissions are not the only mechanism available for app developers to expose (or access) features and components to (or from) other apps. 
Android apps can also interact with each other using \emph{intents}, a
high-level communication abstraction~\cite{androidintents}.
\claim{An app may expose its component(s) to external apps by declaring \texttt{android:exported=true} in the manifest without protecting the component 
with any additional measure, or by adding one or more intent-filters to its declaration in the manifest; exposing it to a type of attack known in 
the literature as a confused deputy attack ~\cite{felt2011permission}}. If the \texttt{exported} attribute is used, it can be protected by adding a 
permission to the component, be it a custom permission or an AOSP one, through checking the caller app's permissions programmatically in the component's 
Java class.

We sought to identify \claim{potentially careless development practices} 
that may lead to components getting exposed without any additional protection.
Exporting components can lead to: $i)$ 
\claim{harmful
or malicious} apps launching an exposed activity, tricking users into 
believing that they are interacting with the benign one; $ii)$ initiating and
binding to unprotected services; and $iii)$ malicious 
apps gaining access to sensitive data or the ability to modify the app's internal state.

We found 6,849 pre-installed 
apps that potentially expose at least one activity in devices from 166 vendors 
and signed by 261 developer signatures 
with \texttt{exported=true}. For services, 4,591 apps (present in 157 vendors) signed by 183 developers including manufacturers, potentially 
exposed one or more of their services to external apps. The top-10 vendors in our dataset account for over 70\% of the potentially
exposed activities and services. 

Other relevant examples include an app that potentially
exposes several activities related to system configurations (device
administration, networking, \etc{}), hence
allowing a malicious developer could access or even tamper a users' device settings. The core package 
\texttt{com.android.mms} found in customized firmware versions across several vendors
also expose services to read WAP messages to other apps.
We also found 8 different instances of a third-party app, 
found in handsets built by two large Android manufacturers, whose intended purpose is 
to provide remote technical support to customers.
This particular service provides remote administration to MNOs,
including the ability to record audio and video, browse files, access system settings, 
and upload/download files. The key service to do so is exposed and can be misused by other apps.

We leave the detailed study of apps vulnerable to confused deputy attacks
and 
the study of the access to these resources by apps 
publicly available on Google Play for future work.

\vspace*{-.5em}
\section{Behavioral Analysis}
\label{sec:appAnalysis}

We analyze the apps in our dataset to identify potentially \claim{harmful} and unwanted behaviors. 
To do this, we leverage both static and dynamic analysis tools to elicit behavior and characterize 
purpose and means. This section describes our analysis pipeline and evidence
of potentially \claim{harmful} and \claim{privacy-intrusive} pre-installed packages.

\vspace*{-.8em}
\subsection{Static Analysis}
\label{sec:appAnalysis:static}

We triage all apps to determine the presence of potentially \claim{harmful}
behaviors. This step allows 
us to obtain a high-level overview of behaviors across the dataset and also provides us with the 
basis to score apps and flag those potentially more interesting. This step is critical since we could
only afford to manually inspect a limited subset of all available apps.

\noindent\textbf{Toolkit.}
Our analysis pipeline integrates various static analysis tools to elicit behavior in Android apps, 
including Androwarn~\cite{androwarn}, FlowDroid~\cite{arzt2014flowdroid}, and Amandroid~\cite{amandroid}, 
as well as a number of custom scripts based on the Apktool~\cite{apktool} and Androguard~\cite{androguard} 
frameworks. In this stage we do not use dynamic analysis tools, which prevents us from identifying 
hidden behaviors that rely on dynamic code uploading (DEX loading) or reflection. This means that 
our results present a lower-bound estimation of all the possible 
potentially \claim{harmful} behaviors. 
We search for apps using DEX loading and reflection to identify targets that deserve manual inspection.

\noindent\textbf{Dataset.}
Because of scalability limitations 
-- our dataset comprises 82,501 APK files with 6,496 unique
package names -- we randomly select one 
APK file for each package name and analyze the resulting set of apps, obtaining an analysis report 
for 48\% of them.
The majority of the remaining packages could not be analyzed due to the absence of a \texttt{classes.dex} 
for odexed files.
Even though in some cases we had the corresponding \texttt{.odex} file, we generally could not deodex 
it since the device's Android framework file was needed to complete this step but \appname{} did not 
collect it.  
Moreover, we could not analyze a small subset of apps due to the limitations of our tools, including errors 
generated during analysis, file size limitations, or analysis tools becoming unresponsive after hours
of processing. Instead, we focused our analysis on the subset of apps for 
which we could generate reports.

\noindent\textbf{Results.}
We processed the analysis reports and identified the presence of the 36
\claim{potentially privacy intrusive behaviors or potentially harmful behaviors 
listed in 
Table}~\ref{table:maliciousIndicators}.
The results suggest that a significant fraction of the analyzed apps could
access and \claim{disseminate} both user and device identifiers, user's
location, and device current configuration. 
\claim{According to our flow analysis, these results  
give the impression that personal data collection and dissemination 
(regardless of the purpose or consent) is not only pervasive 
but also comes pre-installed}. Other a priori concerning behaviors include the possible \claim{dissemination} of contacts 
and SMS contents (164 and 74 apps, respectively), sending SMS (29 apps), and
 making phone calls (339 apps).
Even though there are perfectly legitimate use cases 
for these behaviors, they are also prevalent in harmful and potentially unwanted software.
The distribution of the number of \claim{potentially harmful} behaviors per app follows a power-law distribution. 
Around 25\% of the analyzed apps present at least 5 of these behaviors, 
with almost 1\% of the apps showing 20 or more. The bulk of the distribution relates to the collection of telephony
and network identifiers, interaction with the package manager, and logging activities. This provides 
a glimpse of how pervasive user and device fingerprinting is nowadays.

\begin{table}[t!]
\centering
\begin{adjustbox}{max width=\textwidth}
\begin{tabular}{clrc}
\toprule
\multicolumn{2}{c}{\textbf{\claim{Accessed PII type / behaviors}}} &
\multicolumn{1}{c}{\textbf{Apps (\#)}} &
\multicolumn{1}{c}{\textbf{Apps (\%)}} \\
\midrule
\multirow{11}{2cm}{\centering Telephony identifiers} &
IMEI & 687 & \databar{21.8}\\
& IMSI & 379 & \databar{12}\\
& Phone number & 303 & \databar{9.6}\\
& MCC & 552 & \databar{17.5}\\
& MNC & 552 & \databar{17.5}\\
& Operator name & 315 & \databar{10}\\
& SIM Serial number & 181 & \databar{5.7}\\
& SIM State & 383 & \databar{12.1}\\
& Current country & 194 & \databar{6.2}\\
& SIM country & 196 & \databar{6.2}\\
& Voicemail number & 29 & \databar{0.9}\\
\midrule
\multirow{5}{2cm}{\centering Device settings} &
Software version & 25 & \databar{0.8}\\
& Phone state & 265 & \databar{8.4}\\
& Installed apps & 1,286 & \databar{40.8}\\
& Phone type & 375 & \databar{11.9}\\
& Logs & 2,568 & \databar{81.4}\\
\midrule
\multirow{4}{2cm}{\centering Location} &
GPS & 54 & \databar{1.7}\\
& Cell location & 158 & \databar{5}\\
& CID & 162 & \databar{5.1}\\
& LAC & 137 & \databar{4.3}\\
\midrule
\multirow{5}{2cm}{\centering Network interfaces} &
Wi-Fi configuration & 9 & \databar{0.3}\\
& Current network & 1,373 & \databar{43.5}\\
& Data plan & 699 & \databar{22.2}\\
& Connection state & 71 & \databar{2.3}\\
& Network type & 345 & \databar{10.9}\\
\midrule
\multirow{2}{2cm}{\centering Personal data} &
Contacts & 164 & \databar{11}\\
& SMS & 73 & \databar{2.31}\\
\midrule
\multirow{4}{2cm}{\centering Phone service abuse} &
SMS sending & 29 & \databar{0.92}\\
& SMS interception & 0 & \databar{0}\\
& Disabling SMS notif. & 0 & \databar{0}\\
& Phone calls & 339 & \databar{10.7}\\
\midrule
\multirow{2}{2cm}{\centering Audio/video interception} &
Audio recording & 74 & \databar{2.4}\\
& Video capture & 21 & \databar{0.7}\\
\midrule
\multirow{2}{2cm}{\centering Arbitrary code execution} &
Native code & 775 & \databar{24.6}\\
& Linux commands & 563 & \databar{17.9}\\
\midrule
\multirow{1}{2cm}{\centering Remote conn.} &
Remote connection & 89 & \databar{2.8}\\
\bottomrule
\end{tabular}
\end{adjustbox}
\caption{Volume of apps accessing /\xspace reading PII or  
showing potentially harmful  
behaviors.
The percentage is referred to the subset of triaged packages ($N=3,154$).}
\label{table:maliciousIndicators}
\vspace*{-2em}%
\end{table}

\vspace*{-1em}
\subsection{Traffic Analysis}
\label{sec:appAnalysis:traffic}
While static analysis can be helpful to determine a lower bound of what an app
is capable of, relying on this technique alone gives an incomplete picture of
the real-world behavior of an app. This might be due to code paths that are not 
available at the time of analysis, including those that are within statically- 
and dynamically-linked libraries that are not provided with apps, behaviors 
determined by server-side logic (e.g., due to real-time ad-bidding), 
or code that is loaded at runtime using Java's 
reflection APIs. This limitation of static approaches is generally addressed by 
complementing static analysis with dynamic analysis tools. However, due to various limitations 
(including missing hardware features and software components) it was unfeasible 
for us to run all the pre-installed apps in our dataset in an analysis sandbox. 
Instead, we decided to use the crowd-sourced
\lumenname{} mobile traffic dataset to find  
\claim{evidence of dissemination of personal data}  
from the pre-installed apps by
examining packages that exist
in both datasets. 

\noindent\textbf{Results.}
Of the \appsWithInternet{} pre-installed apps with Internet access permissions,
\appsInLumenFlows{} have at least one flow in the \lumenname{} dataset. At this
point, our analysis of these apps focused on two main aspects: uncovering the
ecosystem of organizations who own the domains that these apps connect to, and
analyzing the types of private information they \claim{could} disseminate from user devices.
To understand the ecosystem of data collection by pre-installed apps, we
studied where the data that is collected by these apps makes its first stop.
We use the Fully-Qualified Domain Names (FQDN) of the servers that are
contacted and use the web crawling and text mining techniques described
in our previous work~\cite{razaghpanah2018apps} to determine the parent
organization who own these domains.

\noindent\textbf{The Big Players.}
Table~\ref{tab:parents_of_domains_in_flows} shows the parent organizations who
own the most popular domains contacted by pre-installed apps in the
\lumenname{} dataset. Of the \domainsFromPreinstalledInLumen{} domains contacted by apps, 
\domainsATSFromPreinstalledInLumen{} belong to well-known Advertising and Tracking
Services (ATS)~\cite{razaghpanah2018apps}. 
These services are represented by organizations like Alphabet,
Facebook, Verizon (now owner of Yahoo!, AOL, and Flurry), Twitter (MoPub's parent
organization), AppsFlyer, comScore, and others.
As expected, Alphabet, the entity that owns and maintains the Android 
platform and many of the largest advertising and tracking services (ATS) in the 
mobile ecosystem~\cite{razaghpanah2018apps},
also owns most of the domains to which pre-installed apps connect to. Moreover, 
vendors who ship their devices with the Google Play Store have to go
through Google's certification program which, in part, entails pre-loading Google's 
services. Among these services is Google's own \texttt{com.google.backuptransport} 
package, which sends a variety of information about the user and the device on
which it runs to Google's servers.

Traffic analysis also confirms that Facebook and Twitter services come pre-installed 
on many phones and are integrated into various apps. 
Many devices also pre-install weather apps like AccuWeather and The Weather Channel.
As reported by previous research efforts, these weather providers
also gather information about the devices and 
their users~\cite{ren2018bug,razaghpanah2018apps}.

\begin{table}[t!]
  \centering
  \scalebox{0.9}{%
    \begin{tabular}{lrr}
      \toprule
      \multicolumn{1}{c}{\textbf{Organization}} & \multicolumn{1}{c}{\textbf{\# of apps}} &
      \multicolumn{1}{c}{\specialcell{\textbf{\# of domains}}} \\
      \midrule
      Alphabet & 566 & 17052 \\
      Facebook & 322 & 3325 \\
      Amazon & 201 & 991 \\
      Verizon Communications & 171 & 320 \\
      Twitter & 137 & 101 \\
      Microsoft & 136 & 408 \\
      Adobe & 116 & 302 \\
      AppsFlyer & 98 & 10 \\
      comScore & 86 & 8 \\
      AccuWeather & 86 & 15 \\
      MoatInc. & 79 & 20 \\
      Appnexus & 79 & 35 \\
      Baidu & 72 & 69 \\
      Criteo & 70 & 62 \\ 
      PerfectPrivacy & 68 & 28 \\ \midrule
      Other ATS & 221 & 362 \\
      \bottomrule
    \end{tabular}
  }\captionof{table}{Top 15 parent ATS organizations by number of apps
connecting to all their associated domains.}
\label{tab:parents_of_domains_in_flows}
\vspace*{-1.5em}
\end{table}

\vspace*{-.5em}
\subsection{Manual Analysis: Relevant Cases}
\label{sec:appAnalysis:manual}
We used the output provided by our static and dynamic analysis pipeline to
score apps and thus flag a reduced subset of packages to inspect manually.
Our goal here was to \claim{confidently} identify \claim{potentially harmful and unwanted}
behavior in pre-installed apps. 
Other apps were added to this set based on the results of our
third-party library and permission analysis performed in Sections~\ref{sec:ecosystem} 
and~\ref{sec:permissions}, respectively.
We manually analyzed 158 apps using standard tools that include DEX disassemblers 
(baksmali), dex-to-java decompilers (jadx, dex2jar), resource analysis tools 
(Apktool), instrumentation tools (Frida), and reverse engineering frameworks 
(radare2 and IDA Pro) for native code analysis. Our main findings can be loosely 
grouped into three large categories: 1) known malware; 2) potential personal data
access and dissemination; 
and 3) potentially harmful apps. Table~\ref{table:manualAnalysis:cases}
provides some examples of the type of behaviors that we found.

\begin{table*}[th!]
\centering
\begin{adjustbox}{max width=\textwidth}
\begin{tabular}{lp{16cm}}
\toprule
\textbf{Family} &
\textbf{Potential Behavior and Prevalence} \\

\midrule
	\multicolumn{2}{l}{\textbf{Known Malware}} \\ \midrule
Triada &
Disseminates PII and other sensitive data (SMS, call logs, contact data,
stored pictures and videos). Downloads additional stages. Roots the device
to install additional apps. \\[1ex]

Rootnik~\cite{rootnik}
& Gains root access to the device. Leaks PII and installs
additional apps. Uses anti-analysis and anti-debugging techniques.\\[1ex]

GMobi~\cite{gmobiDrWeb,upstreamSystems}
& Gmobi Trade Service. Leaks PII, including device serial number and MAC
address, geolocation, installed packages and emails. Receives commands
from servers to (1) send an SMS to a given number; (2) download and
install apps; (3) visit a link; or (4) display a pop-up. It has 
been identified in low-end devices.\\

\midrule
	\multicolumn{2}{l}{\textbf{Potentially Dangerous Apps}} \\ \midrule
Rooting app
& Exposes an unprotected receiver that roots the device upon receiving
a telephony secret code (via intent or dialing \texttt{*\#*\#9527\#*\#*}).\\[1ex]

Blocker
& If the device does not contain a signed file in a particular location,
it loads and enforces 2 blacklists: one containing 103 packages associated with
benchmarking apps, and another with 56 web domains related to phone reviews.\\

\midrule

\multicolumn{2}{l}{\textbf{Potential Personal Data
Access and Dissemination}} \\ \midrule
TrueCaller
& Sends PII to its own servers and embedded third-party ATSes such as AppsFlyer,
Twitter-owned MoPub, Crashlytics, inMobi, Facebook, and others. Uploads phone
call data to at least one of its own domains. \\[1ex]

MetroName ID
& Disseminates PII to its own servers and also to 
third-party services like Piano, a media audience and engagement analytics service that tracks
user's installation of news apps and other partners including those made by
CNBC, Bloomberg, TechCrunch, and The Economist, among others,
the presence of which it reports to its own domains.\\[1ex]

Adups~\cite{kryptowireADUPSfollowup}
& FOTA app. Collects and shares private and PII with their own servers and those
of embedded third-party ATS domains, including Advmob and Nexage. Found worldwide
in 55 brands. \\[1ex]

Stats/Meteor
& Redstone's FOTA service. Uses dynamic code uploading and reflection to
deploy components located in 2 encrypted DEX files. Disseminates around 50 data items
that fully characterize the hardware, the telephony service, the network,
geolocation, and installed packages. Performs behavioral and performance
profiling, including counts of SMS/MMS, calls logs, bytes sent
and transmitted, and usage stats and performance counters on a package-basis.
Silently installs packages on the device and reports what packages are
installed / removed by the user.\\

\bottomrule
\end{tabular}
\end{adjustbox}
\caption{Examples of relevant cases and their potential behaviors found after
manual analysis of a subset of apps.
When referring to personal data dissemination, the term PII encompasses items
enumerated in Table~\ref{table:maliciousIndicators}.}
\label{table:manualAnalysis:cases}
\vspace*{-2em}
\end{table*}

\noindent\textbf{Known Malware.}
We came across \claim{various} isolated instances of known-malware in
the system partition, mostly in
low-end devices but also in some high-end phones. We identified variants of
well-known Android malware families that have been prevalent in the last
few years, including Triada, Rootnik, SnowFox, Xinyin, Ztorg, Iop, and
dubious software developed by GMobi.
\claim{We used VirusTotal to label these samples.}
\claim{According to existing AV reports,} the range of behaviors that such samples
exhibit encompass banking fraud, sending
SMS to premium numbers or subscribing to services, silently installing
additional apps, visiting links, and showing ads, among others. While our method
does not allow us to distinguish whether potentially malicious 
apps are indeed pre-installed or took advantage of system
vulnerabilities to install themselves in the system partition, it is important to
highlight that the presence of pre-installed malware in Android devices 
has been previously reported by various sources
\cite{triadaLow,LokiPreinstalled,upstreamSystems}.
Some of the found samples use Command and Control (C2)
servers still in operation
at the time of this writing.\newline
\noindent\textbf{Personal Data Access and Potential Dissemination.}
Nearly all apps which we identified as able to access PII, 
appear to disseminate it to third-party servers.
We also observed instances of apps with capabilities to perform 
hardware and network fingerprinting, often collected under the term
``device capability,'' and even analytics services that track the installation
and removal 
of apps (notably news apps, such as those made by CNBC, 
Bloomberg, TechCrunch, and The Economist, among others). 
More intrusive behaviors include apps \claim{able to collect and send email} and phone call 
metadata. The most extreme case we analyzed is a \claim{data collection service contained in a 
FOTA service associated with Redstone Sunshine Technology Co., Ltd.}~\cite{redstone}, 
an OTA provider that ``supports 550 million phone users and IoT partners in 40 
countries''~\cite{redstonePressRelease}. This app includes a service that
\claim{can collect 
and disseminate} dozens of data items, including both user and device identifiers, 
behavioral information (counts of SMS and calls sent and received, and statistics about 
network flows) and usage statistics and performance information per installed package.
Overall, this software seems to implement an analytics program that \claim{admits} several monetization 
strategies, from optimized ad targeting to providing performance feedback to both developers 
and manufacturers. We emphasize that the data collected is not only remarkably
extensive and multi-dimensional, 
but also very far away from being anonymous as it is linked to both user and
device IDs.

\noindent\textbf{Potentially dangerous apps.}
We found 612 pre-installed apps that potentially implement engineering- or factory-mode functions
according to their package and app names.
Such functions include relatively harmless tasks, such as hardware tests, but also potentially 
dangerous functions such as the ability to root the device. We found instances of such apps in 
which the rooting function was unprotected in their manifest (\ie{} the component was available 
for every other app to use). We also identified \claim{well-known} 
vulnerable engineering mode apps such like MTKLogger~\cite{johnson17bhus}. Such apps expose unprotected components 
that can be misused by other apps co-located in the device.
\claim{Other examples include a well known manufacturer's 
service, which under certain conditions blacklists connections to a pre-defined list of 56 web 
domains (mobile device review and benchmarking websites, mostly) and disables any installed package 
that matches one of a list of 103 benchmarking apps}.

\vspace*{-.5em}
\section{Study Limitations}
\label{sec:limitations}

\noindent \textbf{Completeness and coverage.}
Our dataset is not complete in terms of Android vendors and models, even though we
cover those with a larger market share, both in the high- and low-end parts
of the spectrum. Our data collection process is also best-effort. 
The lack of background knowledge and documentation required 
performing a detailed case-by-case study and a significant amount of manual inspection. 
In terms of analyzed apps, determining the coverage of our study is difficult since 
we do not know the total number of pre-installed apps in all shipped handsets.

\noindent\textbf{Attribution.}
There is currently no reliable way to accurately find the legitimate developer
of a given pre-installed app by its self-signed signature. 
We have found instances of certificates with just a country code in the \texttt{Issuer}
field, and others with strings suggesting major vendors (\eg Google) signed the app, where 
the apps certainly were not signed by them. The same applies to package and permission
names, many of which are opaque and not named following best-practices. Likewise, the lack of documentation 
regarding custom permissions prevented us from automatizing our analysis. Moreover, a deeper 
study of this issue would require checking whether those permissions are granted in runtime,
tracing the code to fully identify their purpose, and finding whether they are actually used 
by other apps in the wild, and at scale.

\noindent\textbf{Package Manager.}
We do not collect the \texttt{packages.xml} file from our users' devices as it
contains information about all installed packages, and not just pre-installed ones.
We consider that collecting this file would be invasive. This, however, limits
our ability to see if user-installed apps are using services exposed
by pre-installed apps via intents or custom permissions. We tried to
compensate for that with a manual search for public apps that use pre-installed
custom permissions, as discussed in Section~\ref{sec:apiexposing}.

\noindent\textbf{Behavioral coverage.}
Our study mainly relies on static analysis of the samples harvested through
\appname{}, and we only applied dynamic analysis to a selected subset of
\appsInLumenFlows{} packages.
This prevents us from eliciting behaviors that are only available at runtime 
because of the use of code loading and reflection, and also code downloading 
from third-party servers. Despite this, our analysis pipeline served to identify a 
considerable amount of \claim{potentially harmful behaviors}. 
A deeper and broader analysis would possibly uncover more cases.

\noindent \textbf{Identifying rooted devices.}
There is no sure way of knowing whether a device is rooted or not. While
our conservative approach limits the number of false negatives, we have found
occurrences of devices with well-known custom ROMs that were not flagged as
rooted by RootBeer. Moreover, we have found some apps that allow a
third party to root the device on-the-fly to, for example, install new apps on the
system partition as discussed in Section~\ref{sec:appAnalysis:manual}.
Some of these apps can then un-root the phone to avoid detection.
Under the presence 
of such an app on a device, we cannot know for sure if a given package --
particularly a potentially malicious app -- was
pre-installed by an actor in the supply chain, or was installed afterwards.

\vspace*{-.75em}
\section{Related work}
\label{sec:relatedwork}

\noindent\textbf{Android images customization.}
Previous work has been focused on studying modifications made to AOSP images, whether
by adding root certificates~\cite{rootstores}, customizing the default 
apps~\cite{aafer2016harvesting}, or the OS itself~\cite{zhou2014peril}.
In~\cite{aafer2015hare}, Aafer~\etal{} introduced a new class of vulnerability caused 
by the firmware customization process. If an app is removed but a reference to it remains 
in the OS, a malicious app could potentially impersonate it which could lead to privacy 
and security issues. While these studies have focused on Android images as a whole rather
than pre-installed apps, they all show the complexity of the Android ecosystem and underline 
the lack of control over the supply chain.

\noindent\textbf{Android permissions.}
Previous studies on Android permissions have mainly leveraged static analysis
techniques to infer the role of a given permission~\cite{au2012pscout,felt2011android}. 
These studies, however, do not cover newer versions of Android~\cite{Zhauniarovich16Permissions}, 
or custom permissions. In~\cite{jiang2013detecting}, Jiang~\etal{} demonstrated 
how custom permissions are used to expose and protect services. Our work complements 
this study by showing how device makers and third parties alike declare and use 
custom permissions, and make the first step towards a complete and in-depth analysis 
of the whole custom permissions' landscape.

\noindent\textbf{Vulnerabilities in pre-installed apps.}
A recent paper by Wu~\etal{}~\cite{wu2019openports} also used crowdsourcing
mechanisms to detect apps that listen to a given TCP or UDP port and analyze
the vulnerabilities that are caused by this practice. While their study is not
limited to user-installed apps, they show evidence of pre-installed apps
exhibiting this behavior.

\vspace*{-1em}
\section{Discussion and Conclusions}
\label{sec:discussion}
This paper studied, at scale, the vast and unexplored ecosystem of pre-installed 
Android software and its \claim{potential} impact on consumers. This study has made clear that, thanks 
in large part to the open-source nature of the Android platform and the
complexity of its supply chain, organizations of 
various kinds and sizes have the ability to embed their software in custom Android 
firmware versions.
As we demonstrated in this paper, this situation has become a peril to users' privacy 
and even security due to an \claim{abuse} of privilege
or as a result of poor software engineering practices that 
introduce vulnerabilities and dangerous backdoors.

\noindent \textbf{The Supply Chain.}
The myriad of actors involved in the development of pre-installed software and
the supply chain 
range from hardware manufacturers to MNOs and third-party advertising 
and tracking services. These actors have privileged access to system resources   
through their presence in pre-installed apps but also as third-party libraries
embedded in them.
\claim{Potential partnerships and deals -- made behind closed doors between stakeholders 
-- may have made user data a commodity
before users purchase their devices or decide to install software of their own}. 

\noindent \textbf{Attribution.} 
Unfortunately, due to a lack of central authority or trust system to allow verification 
and attribution of the self-signed 
certificates that are used to sign apps, and due to a lack of any 
mechanism to identify the purpose and legitimacy of many of these apps and custom permissions, 
it is difficult to attribute unwanted and harmful 
app behaviors to the party or parties responsible. 
This has broader negative implications for accountability and liability in this ecosystem 
as a whole.

\noindent \textbf{The Role of Users and Informed Consent.}
In the meantime regular Android users are, by and large, unaware of the presence 
of most of the software that comes pre-installed on their Android devices and
their associated privacy risks. Users are clueless about 
\claim{the various data-sharing relationships and partnerships that exist}
between 
companies that have a hand in deciding what comes pre-installed on their phones.
Users' activities, personal data, and habits may be constantly monitored by stakeholders 
that many users may have never heard of, 
let alone consented to collect their data. We have demonstrated instances of devices being 
backdoored by companies with the ability to root and remotely control devices without user 
awareness, and install apps through targeted monetization and user-acquisition campaigns.
Even if users decide to stop or delete some of these apps, they will not be able to do so 
since many of them are core Android services and others 
cannot be permanently removed by the user without root privileges.
It is unclear if the users have actually consented to these practices, or if they were informed 
about them before using the devices (\ie on first boot) in the first place. To clarify this, we 
acquired 6 popular brand-new Android devices from vendors including Nokia, Sony, LG, and Huawei 
from a large Spanish retailer. When booting them, 3 devices did not present a privacy policy at
all, only the Android terms of service.
The rest rendered a privacy policy that only mentions that they collect data about the user, including 
PII such as the IMEI for added value services. 
Note that users have no choice but to accept Android's terms of service, as well as 
the manufacturer's one if presented to the user. Otherwise Android will simply stop booting, 
which will effectively make the device unusable.

\noindent \textbf{Consumer Protection Regulations.} 
While some jurisdictions have very few regulations governing online 
tracking and data collection, there have been a number of movements to regulate and control 
these practices, such as the GDPR in the EU~\cite{gdpr}, 
and California's CCPA~\cite{ccpa} in the US. 
While these efforts are certainly helpful in regulating the rampant invasion of users' privacy 
in the mobile world, they have a long way to go. Most mobile devices
\claim{still lack} a clear and meaningful mechanism to obtain 
informed consent, which is a potential
violation of the GDPR. In fact, it is possible that 
many of the ATSes that come pre-installed on Android devices may not be COPPA-compliant~\cite{reyes2018won}
-- a US federal rule to protect minors from unlawful online tracking~\cite{coppa} --, 
despite the fact that many minors in the US use mobile devices with
pre-installed software that \claim{engage} 
in data collection. This indicates that even in jurisdictions with strict privacy and consumer
protection laws, 
there still remains a large gap between \claim{what is done in practice} and the 
enforcement capabilities of the agencies appointed to uphold the law.

\noindent \textbf{Recommendations.}
To address the issues mentioned above and to make the ecosystem more transparent
we propose a number of recommendations
which are made under the assumption that stakeholders 
are willing to self-regulate and to enhance the status quo.
We are aware that some of these suggestions may inevitably not align with corporate
interests of every organizations in the supply chain, and that an
independent third party 
may be needed to audit the process. Google
might be a prime candidate for it given its capacity for 
licensing vendors and its certification programs.
Alternatively, in absence of self-regulation, governments and regulatory
bodies could step in and enact regulations and execute enforcement actions 
that wrest back some of the control from 
the various actors in the supply chain. We also propose a number of actions that would 
help independent investigators to detect deceptive and potentially \claim{harmful} behaviors.

\noindent {$\bullet$ \textit{Attribution and accountability:}} To combat the difficulty
in attribution and the resulting lack of accountability, we propose the introduction
and use of certificates that are signed by globally-trusted certificate
authorities. Alternatively, it may be possible to build 
a certificate transparency repository dedicated to providing details and attribution
for self-signed certificates used to sign various Android apps, including
pre-installed ones.

\noindent {$\bullet$  \textit{Accessible documentation and consent forms:}} 
Similar to the manner in which
open-source components of Android require any modified version of the code to be made
publicly-available, Android devices can be required to document the specific set of 
apps that have pre-installed, along with their purpose and the entity responsible for each 
piece of software, in a manner that is accessible and understandable to the users. This 
will ensure that at least a reference point exists for users (and regulators) to find accurate information 
about pre-installed apps and their practices. 
Moreover, the results of our small-scale survey of consent forms of some Android 
vendors leaves a lot to 
be desired from a transparency perspective: 
users are not clearly informed about third-party software that is installed on 
their devices, including embedded third-party tracking and advertising services, 
the types of data they collect from them by default, 
and the partnerships that allow 
personal data to be shared over the Internet. 
This necessitates a new form of privacy 
policy suitable for pre-installed apps to be defined (and enforced) to ensure that 
such practices are at least communicated to the user in a clear and accessible
way. This should be accompanied 
by mechanisms to enable users to make informed decisions about how or whether to use such 
devices without having to root them.

\noindent \textbf{Final Remarks.}
Despite a full year of efforts, we were only able to scratch the surface of a much larger 
problem. This work is therefore exploratory, and we hope it will bring more attention to 
the Android supply chain ecosystem and its impact on users' privacy and security. 
We have discussed our results with Google which gave us useful feedback.
Our work was also the basis of a report produced by the Spanish Data
Protection Agency (AEPD)~\cite{aepdnote}.
We will also improve the capabilities and features of both \appname and \lumenname to address 
some of the aforementioned limitations and develop methods to perform dynamic analysis 
of pre-installed software. Given the scale of the ecosystem and the need for manual 
inspections, we will gradually make our dataset (which keeps growing at the time of this 
writing) available to the research community and regulators to aid in future investigations 
and to encourage more research in this area.

\vspace*{-.75em}
\section*{Acknowledgments}

We are deeply grateful to our Firmware Scanner users for
enabling this study, and ElevenPaths for their initial support in this project.
We thank the anonymous reviewers
for their helpful feedback. This project is partially funded by the US 
National Science Foundation (grant CNS-1564329),
the European Union's Horizon 2020 Innovation Action program 
(grant Agreement No. 786741, SMOOTH
Project), the Spanish Ministry of Science, Innovation and Universities
(grants DiscoEdge TIN2017-88749-R and SMOG-DEV TIN2016-79095-C2-2-R),
and the Comunidad de Madrid (grant 
EdgeData-CM P2018/TCS-4499). Any opinions,
findings, conclusions, or recommendations expressed in this
paper are those of the authors and do not reflect the views of
the funding bodies.

{\normalsize \bibliographystyle{acm}
\balance
\bibliography{paper,relatedwork}}

\begin{thebibliography}{10}

\bibitem{adguardMeizu}
{AdGuard - Meizu Incompatibilities}.
\newblock \url{https://github.com/AdguardTeam/AdguardForAndroid/issues/800}.
\newblock [Online; accessed 31-March-2019].

\bibitem{amazonBluPhones}
{Amazon suspends sales of Blu phones for including preloaded spyware, again}.
\newblock
  \url{https://www.theverge.com/2017/7/31/16072786/amazon-blu-suspended-android-spyware-user-data-theft}.
\newblock [Online; accessed 31-March-2019].

\bibitem{aepdnote}
{An\'alisis del software preinstalado en dispositivos Android y riesgos para la
  privacidad de los usuarios}.
\newblock \url{https://www.aepd.es/prensa/2019-03-18.html}.
\newblock [Online; accessed 31-March-2019].

\bibitem{androguard}
{Androguard}.
\newblock \url{https://github.com/androguard/androguard/}.
\newblock [Online; accessed 31-March-2019].

\bibitem{androidcertification}
{Android --- Certified}.
\newblock \url{https://www.android.com/certified/}.
\newblock [Online; accessed 31-March-2019].

\bibitem{LokiPreinstalled}
{Android Adware and Ransomware Found Preinstalled on High-End Smartphones}.
\newblock
  \url{https://www.bleepingcomputer.com/news/security/android-adware-and-ransomware-found-preinstalled-on-high-end-smartphones/}.
\newblock [Online; accessed 31-March-2019].

\bibitem{certifiedpartners}
{Android Certified Partners}.
\newblock \url{https://www.android.com/certified/partners/}.
\newblock [Online; accessed 31-March-2019].

\bibitem{androidcompat}
{Android Compatibility Program Overview}.
\newblock \url{https://source.android.com/compatibility/overview}.
\newblock [Online; accessed 31-March-2019].

\bibitem{androidDev}
{Android Developer Documentation}.
\newblock \url{https://developer.android.com/}.
\newblock [Online; accessed 31-March-2019].

\bibitem{androidmarketshare1}
{Android Trackers}.
\newblock \url{https://fiksu.com/resources/android_trackers/}.
\newblock [Online; accessed 31-March-2019].

\bibitem{gmobiDrWeb}
{Android.Gmobi.1}.
\newblock \url{https://vms.drweb.com/virus/?_is=1&i=7999623&lng=en}.
\newblock [Online; accessed 31-March-2019].

\bibitem{androwarn}
{Androwarn--Yet another static code analyzer for malicious Android
  applications}.
\newblock \url{https://github.com/maaaaz/androwarn}.
\newblock [Online; accessed 31-March-2019].

\bibitem{apktool}
{Apktool--A tool for reverse engineering Android apk files}.
\newblock \url{https://ibotpeaches.github.io/Apktool/}.
\newblock [Online; accessed 31-March-2019].

\bibitem{newley}
{App Traps: How Cheap Smartphones Siphon User Data in DevelopingmCountries}.
\newblock
  \url{https://www.wsj.com/articles/app-traps-how-cheap-smartphones-help-themselves-to-user-data-1530788404}.
\newblock [Online; accessed 31-March-2019].

\bibitem{signedapps}
{Application signing}.
\newblock \url{https://developer.android.com/studio/publish/app-signing}.
\newblock [Online; accessed 31-March-2019].

\bibitem{appseerecords}
{Appsee --- Features}.
\newblock \url{https://www.appsee.com/features}.
\newblock [Online; accessed 31-March-2019].

\bibitem{appsee}
{Appsee Mobile App Analytics}.
\newblock \url{https://www.appsee.com/}.
\newblock [Online; accessed 31-March-2019].

\bibitem{asurion}
{Asurion}.
\newblock \url{https://www.asurion.com/}.
\newblock [Online; accessed 31-March-2019].

\bibitem{baidugeocoding}
{Baidu Geocoding API}.
\newblock \url{http://api.map.baidu.com/lbsapi/geocoding-api.htm}.
\newblock [Online; accessed 31-March-2019].

\bibitem{baidusdk}
{Baidu SDK}.
\newblock \url{https://developer.baidu.com/}.
\newblock [Online; accessed 31-March-2019].

\bibitem{ccpa}
{California Consumer Privacy Act}.
\newblock
  \url{https://leginfo.legislature.ca.gov/faces/billTextClient.xhtml?bill_id=201720180AB375}.
\newblock [Online; accessed 31-March-2019].

\bibitem{redstonePressRelease}
{China Mobile Network Partner Redstone Moves into Robotics}.
\newblock \url{https://www.prweb.com/releases/2017/04/prweb14212503.htm}.
\newblock [Online; accessed 31-March-2019].

\bibitem{coppa}
{COPPA - Children's Online Privacy Protection Act}.
\newblock \url{http://coppa.org/}.
\newblock [Online; accessed 31-March-2019].

\bibitem{huaweiGameSkytoneCVE}
{CVE-2017-2709}.
\newblock \url{https://cve.mitre.org/cgi-bin/cvename.cgi?name=CVE-2017-2709}.
\newblock [Online; accessed 31-March-2019].

\bibitem{samsungAccountCVE}
{CVE-2017-2709}.
\newblock \url{https://cve.mitre.org/cgi-bin/cvename.cgi?name=cve-2015-0864}.
\newblock [Online; accessed 31-March-2019].

\bibitem{customperms}
{Define a Custom Permission}.
\newblock
  \url{https://developer.android.com/guide/topics/permissions/defining}.
\newblock [Online; accessed 31-March-2019].

\bibitem{digitalTurbinePolicy}
{Digital Turbine - Privacy Policy}.
\newblock \url{https://www.digitalturbine.com/privacy-policy/}.
\newblock [Online; accessed 31-March-2019].

\bibitem{estimote}
{Estimote --- indoor location with bluetooth beacons and mesh}.
\newblock \url{https://estimote.com/}.
\newblock [Online; accessed 31-March-2019].

\bibitem{gdpr}
{EU General Data Protection Regulation (GDPR)}.
\newblock \url{https://eugdpr.org/}.
\newblock [Online; accessed 31-March-2019].

\bibitem{huaweieu}
{Europe should be wary of Huawei, EU tech official says}.
\newblock
  \url{https://www.reuters.com/article/us-eu-china-huawei-idUSKBN1O611X}.
\newblock [Online; accessed 31-March-2019].

\bibitem{exusUK}
{EXUS}.
\newblock \url{https://www.exus.co.uk}.
\newblock [Online; accessed 31-March-2019].

\bibitem{facebook}
{Facebook Gave Device Makers Deep Access to Data on Users and Friends}.
\newblock
  \url{https://www.nytimes.com/interactive/2018/06/03/technology/facebook-device-partners-users-friends-data.html}.
\newblock [Online; accessed 31-March-2019].

\bibitem{nytfbinvestigation}
{Facebook’s Data Deals Are Under Criminal Investigation}.
\newblock
  \url{https://www.nytimes.com/2019/03/13/technology/facebook-data-deals-investigation.html}.
\newblock [Online; accessed 31-March-2019].

\bibitem{scannerapp}
{Firmware Scanner}.
\newblock
  \url{https://play.google.com/store/apps/details?id=org.imdea.networks.iag.preinstalleduploader}.
\newblock [Online; accessed 31-March-2019].

\bibitem{androidmarketshare0}
{Global market share held by leading smartphone vendors}.
\newblock
  \url{https://www.statista.com/statistics/271496/global-market-share-held-by-smartphone-vendors-since-4th-quarter-2009/}.
\newblock [Online; accessed 31-March-2019].

\bibitem{gmobi}
{GMobi --- General Mobile Corporation}.
\newblock \url{http://www.generalmobi.com/en/}.
\newblock [Online; accessed 31-March-2019].

\bibitem{gcm_perms}
{Google Cloud Messaging}.
\newblock
  \url{https://developers.google.com/cloud-messaging/android/android-migrate-fcm}.
\newblock [Online; accessed 31-March-2019].

\bibitem{hiya}
{Hiya}.
\newblock \url{https://hiya.com/}.
\newblock [Online; accessed 31-March-2019].

\bibitem{hiyaDataPolicy}
{Hiya Partners}.
\newblock \url{https://hiya.com/hiya-data-policy}.
\newblock [Online; accessed 31-March-2019].

\bibitem{truecallerdata1}
{How does Truecaller get its data?}
\newblock
  \url{https://support.truecaller.com/hc/en-us/articles/212638485-How-does-Truecaller-get-its-data}.
\newblock [Online; accessed 31-March-2019].

\bibitem{infinum}
{Infinum Inc.}
\newblock \url{https://infinum.co}.
\newblock [Online; accessed 31-March-2019].

\bibitem{androidintents}
{Intents and Intent Filters - Android Developers}.
\newblock \url{https://developer.android.com/guide/components/intents-filters}.
\newblock [Online; accessed 31-March-2019].

\bibitem{ironsrc}
{IronSource --- App monetization done right}.
\newblock \url{https://www.ironsrc.com/}.
\newblock [Online; accessed 31-March-2019].

\bibitem{ironSourceAura}
{IronSource - AURA}.
\newblock \url{https://company.ironsrc.com/enterprise-solutions/}.
\newblock [Online; accessed 31-March-2019].

\bibitem{ironSourceAuraSlide}
{IronSource - Aura for Advertisers}.
\newblock \url{https://www.slideshare.net/ironSource/aura-for-advertisers}.
\newblock [Online; accessed 31-March-2019].

\bibitem{kryptowireADUPS}
{Kryptowire Discovers Mobile Phone Firmware that Transmitted Personally
  Identifiable Information (PII) without User Consent or Disclosure}.
\newblock \url{https://www.kryptowire.com/adups_security_analysis.html}.
\newblock [Online; accessed 31-March-2019].

\bibitem{kryptowireADUPSfollowup}
{Kryptowire Provides Technical Details on Black Hat 2017 Presentation: Observed
  ADUPS Data Collection \& Data Transmission}.
\newblock
  \url{https://www.kryptowire.com/observed_adups_data_collection_behavior.html}.
\newblock [Online; accessed 31-March-2019].

\bibitem{locationlabs}
{locationlabs by Avast}.
\newblock \url{https://www.locationlabs.com/}.
\newblock [Online; accessed 31-March-2019].

\bibitem{lumenapp}
{Lumen Privacy Monitor}.
\newblock
  \url{https://play.google.com/store/apps/details?id=edu.berkeley.icsi.haystack}.
\newblock [Online; accessed 31-March-2019].

\bibitem{systemPerms}
{Manifest permissions}.
\newblock
  \url{https://developer.android.com/reference/android/Manifest.permission}.
\newblock [Online; accessed 31-March-2019].

\bibitem{appbrain}
{Monetize, advertise and analyze Android apps}.
\newblock \url{https://www.appbrain.com}.
\newblock [Online; accessed 31-March-2019].

\bibitem{oneplus3}
{OnePlus Device Root Exploit: Backdoor in EngineerMode App for Diagnostics
  Mode}.
\newblock
  \url{https://www.nowsecure.com/blog/2017/11/14/oneplus-device-root-exploit-
  backdoor-engineermode-app-diagnostics-mode/}.
\newblock [Online; accessed 31-March-2019].

\bibitem{oneplus2}
{OnePlus left a backdoor in its devices capable of root access}.
\newblock
  \url{http://www.androidpolice.com/2017/11/15/oneplus-left-backdoor-devices-capable-root-access/}.
\newblock [Online; accessed 31-March-2019].

\bibitem{oneplus1}
{OnePlus OxygenOS built-in analytics}.
\newblock \url{https://www.chrisdcmoore.co.uk/post/oneplus-analytics/}.
\newblock [Online; accessed 31-March-2019].

\bibitem{oneplus}
{OnePlus Secret Backdoor}.
\newblock \url{https://www.theregister.co.uk/2017/11/14/oneplus_backdoor/}.
\newblock [Online; accessed 31-March-2019].

\bibitem{perms}
{Permissions overview}.
\newblock
  \url{https://developer.android.com/guide/topics/permissions/overview.html}.
\newblock [Online; accessed 31-March-2019].

\bibitem{truecaller}
{Phone Number Search --- TrueCaller}.
\newblock \url{https://www.truecaller.com/}.
\newblock [Online; accessed 31-March-2019].

\bibitem{privacygrade}
{Privacy Grade}.
\newblock \url{http://privacygrade.org}.
\newblock [Online; accessed 31-March-2019].

\bibitem{privacystar}
{PrivacyStar}.
\newblock \url{https://privacystar.com}.
\newblock [Online; accessed 31-March-2019].

\bibitem{privacystarpolicy}
{PrivacyStar Privacy Policy}.
\newblock \url{https://privacystar.com/privacy-policy/}.
\newblock [Online; accessed 31-March-2019].

\bibitem{redstone}
{Redstone}.
\newblock \url{http://www.redstone.net.cn/}.
\newblock [Online; accessed 31-March-2019].

\bibitem{rootnik}
{Rootnik Android Trojan Abuses Commercial Rooting Tool and Steals Private
  Information}.
\newblock
  \url{https://unit42.paloaltonetworks.com/rootnik-android-trojan-abuses-commercial-rooting-tool-and-steals-private-information/}.
\newblock [Online; accessed 31-March-2019].

\bibitem{rootbeer}
{Simple to use root checking Android library}.
\newblock \url{https://github.com/scottyab/rootbeer}.
\newblock [Online; accessed 31-March-2019].

\bibitem{smaatoGeo}
{Smaato Blog}.
\newblock
  \url{https://blog.smaato.com/everything-you-need-to-know-about-location-based-mobile-advertising}.
\newblock [Online; accessed 31-March-2019].

\bibitem{synchronossprivacypolicy}
{Synchronoss Technologies - Privacy Policy}.
\newblock \url{https://synchronoss.com/privacy-policy/#datacollected}.
\newblock [Online; accessed 31-March-2019].

\bibitem{triadaLow}
{Triada Trojan Found in Firmware of Low-Cost Android Smartphones}.
\newblock
  \url{https://www.bleepingcomputer.com/news/security/android-adware-and-ransomware-found-preinstalled-on-high-end-smartphones/}.
\newblock [Online; accessed 31-March-2019].

\bibitem{upstreamSystems}
{Upstream - Low-end Android smartphones sold with pre-installed malicious
  software in emerging markets}.
\newblock
  \url{https://www.upstreamsystems.com/pre-installed-malware-android-smartphones/
  }.
\newblock [Online; accessed 31-March-2019].

\bibitem{AndroidVPNService}
{VPN Service}.
\newblock \url{https://developer.android.com/reference/android/net/VpnService}.
\newblock [Online; accessed 31-March-2019].

\bibitem{androidCentralFacebookAppManager}
{What is ``com,facebook,app manager'' and why is it trying to download
  Instagram, Facebook, and Messenger}.
\newblock
  \url{https://forums.androidcentral.com/android-apps/547447-what-com-facebook-app-manager-why-trying-download-instagram-facebook-messenge.html}.
\newblock [Online; accessed 31-March-2019].

\bibitem{XDAFacebookSystem}
{XDA-Developers Forum (Galaxy Note 4). com.facebook.appmanager}.
\newblock
  \url{https://forum.xda-developers.com/note-4/themes-apps/com-facebook-appmanager-t2919151}.
\newblock [Online; accessed 31-March-2019].

\bibitem{truecallerdata2}
{Your Data Is Our Data: A Truecaller Breakdown}.
\newblock
  \url{https://techcabal.com/2018/05/02/your-data-is-our-data-a-truecaller-breakdown/}.
\newblock [Online; accessed 31-March-2019].

\bibitem{aafer2015hare}
{\sc Aafer, Y., Zhang, N., Zhang, Z., Zhang, X., Chen, K., Wang, X., Zhou, X.,
  Du, W., and Grace, M.}
\newblock {Hare Hunting In The Wild Android: A Study On The Threat Of Hanging
  Attribute References}.
\newblock In {\em {\ccs{}}\/} ({2015}).

\bibitem{aafer2016harvesting}
{\sc Aafer, Y., Zhang, X., and Du, W.}
\newblock {Harvesting Inconsistent Security Configurations In Custom Android
  ROMs Via Differential Analysis.}
\newblock In {\em {\usenix{}}\/} ({2016}).

\bibitem{arzt2014flowdroid}
{\sc Arzt, S., Rasthofer, S., Fritz, C., Bodden, E., Bartel, A., Klein, J.,
  Le~Traon, Y., Octeau, D., and McDaniel, P.}
\newblock {Flowdroid: Precise context, flow, field, object-sensitive and
  lifecycle-aware taint analysis for android apps}.
\newblock {\em {\sigplan{}}\/} ({2014}).

\bibitem{au2012pscout}
{\sc Au, K. W.~Y., Zhou, Y.~F., Huang, Z., and Lie, D.}
\newblock {PScout: Analyzing The Android Permission Specification}.
\newblock In {\em {\ccs{}}\/} ({2012}).

\bibitem{menloreport}
{\sc Dittrich, D., and Kenneally, E.}
\newblock {The Menlo Report: Ethical principles guiding information and
  communication technology research}.
\newblock {\em {US Department of Homeland Security}\/} ({2012}).

\bibitem{triadaTrojan}
{\sc {Dr Web}}.
\newblock {Trojan preinstalled on Android devices infects applications’
  processes and downloads malicious modules}.
\newblock \url{http://news.drweb.com/news/?i=11390&lng=en}.
\newblock [Online; accessed 31-March-2019].

\bibitem{felt2011android}
{\sc Felt, A.~P., Chin, E., Hanna, S., Song, D., and Wagner, D.}
\newblock {Android Permissions Demystified}.
\newblock In {\em {\ccs{}}\/} ({2011}).

\bibitem{felt2011permission}
{\sc Felt, A.~P., Wang, H.~J., Moshchuk, A., Hanna, S., and Chin, E.}
\newblock {Permission Re-Delegation: Attacks And Defenses.}
\newblock In {\em {\usenix{}}\/} ({2011}).

\bibitem{ikram2016analysis}
{\sc Ikram, M., Vallina-Rodriguez, N., Seneviratne, S., Kaafar, M.~A., and
  Paxson, V.}
\newblock {An analysis of the privacy and security risks of android vpn
  permission-enabled apps}.
\newblock In {\em {\imc{}}\/} ({2016}).

\bibitem{jiang2013detecting}
{\sc Jiang, Y. Z.~X., and Xuxian, Z.}
\newblock {Detecting Passive Content Leaks And Pollution In Android
  Applications}.
\newblock In {\em {\ndss{}}\/} ({2013}).

\bibitem{johnson17bhus}
{\sc {Johnson, Ryan and Stavrou, Angelos and Benameur, Azzedine}}.
\newblock {All Your SMS \& Contacts Belong to ADUPS \& Others}.
\newblock
  \url{https://www.blackhat.com/docs/us-17/wednesday/us-17-Johnson-All-Your-SMS-&-Contacts-Belong-To-Adups-&-Others.pdf}.
\newblock [Online; accessed 31-March-2019].

\bibitem{li2016investigation}
{\sc Li, L., Bissyand{\'e}, T.~F., Klein, J., and Le~Traon, Y.}
\newblock {An investigation into the use of common libraries in android apps}.
\newblock In {\em {Proceedings of the International Conference on Software
  Analysis, Evolution, and Reengineering (SANER)}\/} ({2016}).

\bibitem{pan2018panoptispy}
{\sc Pan, E., Ren, J., Lindorfer, M., Wilson, C., and Choffnes, D.}
\newblock {Panoptispy: Characterizing Audio and Video Exfiltration from Android
  Applications}.
\newblock {\em {\pets{}} {2018}\/}.

\bibitem{razaghpanah2018apps}
{\sc Razaghpanah, A., Nithyanand, R., Vallina-Rodriguez, N., Sundaresan, S.,
  Allman, M., Kreibich, C., and Gill, P.}
\newblock {Apps, Trackers, Privacy, and Regulators: A Global Study of the
  Mobile Tracking Ecosystem}.
\newblock In {\em {\ndss{}}\/} ({2018}).

\bibitem{lumen}
{\sc Razaghpanah, A., Vallina-Rodriguez, N., Sundaresan, S., Kreibich, C.,
  Gill, P., Allman, M., and Paxson, V.}
\newblock {Haystack: In situ mobile traffic analysis in user space}.
\newblock {\em {arXiv preprint arXiv:1510.01419}\/} ({2015}).

\bibitem{ren2018bug}
{\sc Ren, J., Lindorfer, M., Dubois, D.~J., Rao, A., Choffnes, D., and
  Vallina-Rodriguez, N.}
\newblock {Bug Fixes, Improvements,... and Privacy Leaks}.

\bibitem{reyes2018won}
{\sc Reyes, I., Wijesekera, P., Reardon, J., On, A. E.~B., Razaghpanah, A.,
  Vallina-Rodriguez, N., and Egelman, S.}
\newblock {"Won’t Somebody Think of the Children?" Examining COPPA Compliance
  at Scale}.
\newblock {\em {\pets{}}\/} ({2018}).

\bibitem{rootstores}
{\sc Vallina-Rodriguez, N., Amann, J., Kreibich, C., Weaver, N., and Paxson,
  V.}
\newblock {A Tangled Mass: The Android Root Certificate Stores}.
\newblock In {\em {\conext{}}\/} ({2014}).

\bibitem{vallina2012breaking}
{\sc Vallina-Rodriguez, N., Shah, J., Finamore, A., Grunenberger, Y.,
  Papagiannaki, K., Haddadi, H., and Crowcroft, J.}
\newblock {Breaking for commercials: characterizing mobile advertising}.
\newblock In {\em {\imc{}}\/} ({2012}).

\bibitem{libradarplusplus}
{\sc Wang, H., Liu, Z., Liang, J., Vallina-Rodriguez, N., Guo, Y., Li, L.,
  Tapiador, J., Cao, J., and Xu, G.}
\newblock {Beyond Google Play: A Large-Scale Comparative Study of Chinese
  Android App Markets}.
\newblock In {\em {\imc{}}\/} ({2018}).

\bibitem{amandroid}
{\sc Wei, F., Roy, S., Ou, X., and Robby}.
\newblock {Amandroid: A Precise and General Inter-component Data Flow Analysis
  Framework for Security Vetting of Android Apps}.
\newblock In {\em {\ccs{}}\/} ({2014}).

\bibitem{wu2019openports}
{\sc Wu, D., Gao, D., Chang, R. K.~C., He, E., Cheng, E. K.~T., , and Deng,
  R.~H.}
\newblock {Understanding Open Ports In Android Applications: Discovery,
  Diagnosis, And Security Assessment}.
\newblock {\em {\ndss{}}\/} ({2019}).

\bibitem{Zhauniarovich16Permissions}
{\sc Zhauniarovich, Y., and Gadyatskaya, O.}
\newblock {Small Changes, Big Changes: An Updated View On The Android
  Permission System}.
\newblock In {\em {Research in Attacks, Intrusions, and Defenses}\/} ({2016}).

\bibitem{zhou2014peril}
{\sc Zhou, X., Lee, Y., Zhang, N., Naveed, M., and Wang, X.}
\newblock {The Peril Of Fragmentation: Security Hazards In Android Device
  Driver Customizations}.
\newblock In {\em {\oakland{}}\/} ({2014}).

\end{thebibliography}

\newpage

\appendix

\subsection{Userbase distribution}
\label{sec:userbase}

Table~\ref{tab:geodistrib} describes our userbase geographical distribution.

\begin{center}
  \begin{adjustbox}{max width=.7\linewidth}
    \begin{tabular}{lrrrr}
      \toprule
      \multirow{ 2}{*}{\specialcell{Country\\(N=\locationsall)}} &
      \multirow{ 2}{*}{\specialcell{Samples}} &
      \multicolumn{2}{c}{\specialcell{Vendors}} &
      \multirow{ 2}{*}{\specialcell{Vendor's\\share}} \\ \cmidrule{3-4}
      & & Total & Unique & \\
      \midrule
      USA & \locationus{} & \totalVendorsUS{}  & \uniqueVendorsUS{} & \vendorsShareUS{} \\
      Spain & \locationes{} & \totalVendorsES{}  & \uniqueVendorsES{} & \vendorsShareES{}  \\
      Indonesia & \locationid{} & \totalVendorsID{} & \uniqueVendorsID{} & \vendorsShareID{} \\
      Italy & \locationit{} & \totalVendorsIT{} & \uniqueVendorsGB{} &  \vendorsShareIT{}  \\
      UK & \locationgb{} & \totalVendorsGB{}  & \uniqueVendorsGB{} & \vendorsShareGB{}  \\
      Mexico & \locationmx{} & \totalVendorsMX{}  & \uniqueVendorsMX{} & \vendorsShareMX{}  \\
      Thailand & \locationth{} & \totalVendorsTH{}  & \uniqueVendorsTH{} & \vendorsShareTH{}  \\
      Germany & \locationde{} & \totalVendorsDE{} & \uniqueVendorsDE{} & \vendorsShareDE{}  \\
      Belgium & \locationbe{} & \totalVendorsBE{}  & \uniqueVendorsBE{} & \vendorsShareBE{}  \\
      Netherlands & \locationnl{} & \totalVendorsNL{}  & \uniqueVendorsNL{} &  \vendorsShareNL{} \\
      \midrule
      \textbf{Total countries} & \textbf{\locationsall{}} & \textbf{---} & \textbf{\totalvendors{}} \\
      \bottomrule
    \end{tabular}
  \end{adjustbox}
  \captionof{table}{Geographical distribution of our users. Only the top 10 countries are shown.}\label{tab:geodistrib}
\end{center}

\newpage

\begin{minipage}{\textwidth}
  \subsection{Custom permissions}
  \label{sec:customPermAppendix}

  Table~\ref{table:appendix:perms} reports a subset of custom permissions defined by
  device vendors, MNOs, third-party services, and chipset manufacturers.
  \begin{center}
    \begin{adjustbox}{max width=\textwidth}
      \begin{tabular}{l l l l} 
        \toprule
        \multicolumn{4}{c}{\textbf{MANUFACTURER PERMISSIONS}} \\[.5em]
        \textbf{Package name} & \textbf{Developer Signature} & \textbf{Vendor(s)} & \textbf{Permission} \\ \midrule
        \texttt{com.sonyericsson.facebook.proxylogin} & Sony Ericsson (SE) & Sony 
        & \textit{com.sonyericsson.permission.FACEBOOK} \\
        \texttt{com.sonymobile.twitter.account} & Sony Ericsson (SE) & Sony 
        & \textit{com.sonymobile.permission.TWITTER} \\ 
        \texttt{android} & Sony Ericsson (SE)  & Sony 
        & \textit{com.sonymobile.googleanalyticsproxy.permission.GOOGLE\_ANALYTICS} \\
        \texttt{com.htc.socialnetwork.facebook} & Android (TW) & HTC & 
        \textit{*.permission.SYSTEM\_USE} \\
        \texttt{com.sonymobile.gmailreaderservice} & Sony Ericsson (SE) & Sony & 
        \textit{com.sonymobile.permission.READ\_GMAIL} \\ 
        \texttt{com.sec.android.daemonapp} & Samsung Corporation (KR) & Samsung & 
        \textit{*.ap.accuweather.ACCUWEATHER\_DAEMON\_ACCESS\_PROVIDER} \\ 
        \texttt{android} & Lenovo (CN) & Lenovo 
        & \textit{android.permission.LENOVO\_MDM} \\
        \texttt{com.asus.loguploaderproxy} & AsusTek (TW) & Asus 
        & \textit{asus.permission.MOVELOGS} \\ 
        \texttt{com.miui.core} & Xiaomi (CN) & Xiaomi 
        & \textit{miui.permission.DUMP\_CACHED\_LOG} \\
        \texttt{android} & Samsung (KR) & Samsung 
        & \textit{com.sec.enterprise.knox.KNOX\_GENERIC\_VPN} \\
        \texttt{com.sec.enterprise.permissions} & Samsung (KR) & Samsung 
        & \textit{android.permission.sec.MDM\_ENTERPRISE\_VPN\_SOLUTION} \\
        \texttt{com.android.vpndialogs} & Meizu (CN) & Meizu 
        & \textit{com.meizu.permission.CONTROL\_VPN} \\
        \bottomrule \\[1em]
        \toprule
        \multicolumn{4}{c}{\textbf{MNO PERMISSIONS}} \\[.5em]
        \textbf{Package name} & \textbf{Developer Signature} & \textbf{MNO} & \textbf{Permission} \\ \midrule
        \texttt{com.android.mms} & ZTE & T-Mobile US & 
        \textit{com.tmobile.comm.RECEIVE\_METRICS} \\
        \texttt{com.lge.ipservice} & LG & T-Mobile US 
        & \textit{com.tmobile.comm.RECEIVE\_METRICS} \\
        \texttt{hr.infinum.mojvip} & Infinum (HR)~\cite{infinum} & H1 Croatia 
        & \textit{hr.infinum.mojvip.permission.RECEIVE\_ADM\_MESSAGE}	\\
        \texttt{com.locationlabs.cni.att} & AT\&T (US) & AT\&T (US)~\cite{locationlabs} 
        & \textit{com.locationlabs.cni.att.permission.BROADCAST} \\
        \texttt{com.asurion.android.verizon.vms} & Asurion (US)~\cite{asurion} & Verizon (US) 
        & \textit{com.asurion.android.verizon.vms.permission.C2D\_MESSAGE} \\ 
        \texttt{jp.naver.line.android} & Naver (JP) & South Korea Telekom &
        \textit{com.skt.aom.permission.AOM\_RECEIVE} \\
        \bottomrule\\[1em]
        \toprule
        \multicolumn{4}{c}{\textbf{THIRD-PARTY SERVICE PERMISSIONS}} \\[.5em]
        \textbf{Package name} & \textbf{Developer Signature} & \textbf{Provider} & \textbf{Permission} \\ \midrule
   \texttt{com.facebook.system} & Facebook & Facebook & 
    \textit{*.ACCESS} \\
   \texttt{com.amazon.kindle} & Amazon & Amazon & 
    \textit{com.amazon.identity.auth.device.perm.AUTH\_SDK} \\	
   \texttt{com.huawei.android.totemweather} & Huawei (CN) & Baidu & 
    \textit{android.permission.BAIDU\_LOCATION\_SERVICE} \\	
   \texttt{com.oppo.findmyphone} & Oppo (CN) & Baidu &
    \textit{android.permission.BAIDU\_LOCATION\_SERVICE} \\	
   \texttt{com.dti.sliide} & Logia & Digital Turbine &
    \textit{com.digitalturbine.ignite.ACCESS\_LOG} \\
   \texttt{com.dti.att} & Logia & Digital Turbine & 
    \textit{com.dti.att.permission.APP\_EVENTS} \\
   \texttt{com.ironsource.appcloud.oobe.wiko} & ironSource & ironSource & 
    \textit{com.ironsource.aura.permission.C2D\_MESSAGE} \\
   \texttt{com.vcast.mediamanager} & Verizon (US) & Synchronoss &
    \textit{com.synchronoss.android.sync.provider.FULL\_PERMISSION} \\
  \texttt{com.myvodafone.android} & Vodafone (GR) & Exus &
    \textit{uk.co.exus.permission.C2D\_MESSAGE} \\
  \texttt{com.trendmicro.freetmms.gmobi} & TrendMicro (TW) & GMobi &
    \textit{com.trendmicro.androidmup.ACCESS\_TMMSMU\_REMOTE\_SERVICE} \\
  \texttt{com.skype.rover} & Skype (GB) & Skype  & 
    \textit{com.skype.android.permission.READ\_CONTACTS} \\
  \texttt{com.cleanmaster.sdk} & Samsung (KR) & CleanMaster  &
    \textit{com.cleanmaster.permission.sdk.clean} \\
  \texttt{com.netflix.partner.activation} & Netflix (US) & Netflix &
    \textit{*.permission.CHANNEL\_ID} \\
  \bottomrule \\[1em]
    \toprule
  \multicolumn{4}{c}{\textbf{CHIPSET PERMISSIONS}} \\[.5em]
   \textbf{Package name} & \textbf{Developer Signature} & \textbf{Provider}  & \textbf{Permission} \\ \midrule
  \texttt{com.qualcomm.location} & ZTE (CN) & Qualcomm & 
      \textit{com.qualcomm.permission.IZAT} \\
  \texttt{com.mediatek.mtklogger} & TCL (CN) & MediaTek & 
    \textit{com.permission.MTKLOGGER} \\
  \texttt{com.android.bluetooth} & Samsung (KR) & Broadcom &  
    \textit{broadcom.permission.BLUETOOTH\_MAP} \\
  \bottomrule
\end{tabular}
\end{adjustbox}
  \captionof{table}{Custom permission examples. The wildcard * represents the
    package name whenever the permission prefix and the package name overlap.}
  \label{table:appendix:perms}
\end{center}
\end{minipage}

\end{document}